\documentclass[fleqn,usenatbib]{mnras}

\usepackage[T1]{fontenc}
\usepackage{ae,aecompl}

\usepackage{graphicx}	
\usepackage{amsmath}	
\usepackage{amssymb}	

\usepackage{ulem}
\usepackage{soul}
\usepackage{natbib}

\newcommand{\msun}{$\rm M_\odot$}

\title[Accretion of Galaxy Groups]{Accretion of Galaxy Groups into Galaxy Clusters}

\author[J. A. Benavides, L. V. Sales and M. G. Abadi]{
Jos\'e A. Benavides$^{1}$\href{https://orcid.org/0000-0003-1896-0424}{\includegraphics[scale=0.8]{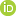}}\thanks{E-mail: jose.astroph@gmail.com},
Laura V. Sales$^{2}$\href{https://orcid.org/0000-0002-3790-720X}{\includegraphics[scale=0.8]{fig_paper_groups_ill_mnras/orcid.png}} and
Mario. G. Abadi$^{1,3}$\href{https://orcid.org/0000-0003-3055-6678}{\includegraphics[scale=0.8]{fig_paper_groups_ill_mnras/orcid.png}}
\\
$^{1}$Instituto de Astronom\'ia Te\'orica y Experimental, CONICET-UNC, Laprida 854, X5000BGR, C\'ordoba, Argentina\\
$^{2}$Department of Physics and Astronomy, University of California, Riverside, CA, 92521, USA\\
$^{3}$Observatorio Astron\'omico de C\'ordoba, Universidad Nacional de C\'odoba, Laprida 854, X5000BGR, C\'ordoba, Argentina\\
}

\date{Accepted XXX. Received YYY; in original form ZZZ}

\pubyear{2020}

\begin{document}
\label{firstpage}
\pagerange{\pageref{firstpage}--\pageref{lastpage}}
\maketitle

\begin{abstract}

  We study the role of group infall in the assembly and dynamics of galaxy clusters in $\Lambda$CDM. We select $10$ clusters with virial mass $M_{\rm 200} \sim 10^{14}$\msun\; from the cosmological hydrodynamical simulation Illustris and follow their galaxies with stellar mass $M_{\star} \geq 1.5 \times 10^8$\msun. A median of $\sim 38\%$ of surviving galaxies at $z=0$ are accreted as part of groups and did not infall directly from the field, albeit with significant cluster-to-cluster scatter. The evolution of these galaxy associations is quick, with observational signatures of their common origin eroding rapidly in $1$-$3$ Gyr after infall. Substructure plays a dominant role in fostering the conditions for galaxy mergers to happen, even within the cluster environment. Integrated over time, we identify (per cluster) an average of $17 \pm 9$ mergers that occur in infalling galaxy associations, of which $7 \pm 3$ occur well within the virial radius of their cluster hosts. The number of mergers shows large dispersion from cluster to cluster, with our most massive system having $42$ mergers above our mass cut-off. These mergers, which are typically gas rich for dwarfs and a combination of gas rich and gas poor for $M_{\star} \sim 10^{11}$\msun, may contribute significantly within $\Lambda$CDM to the formation of specific morphologies, such as lenticulars (S0) and blue compact dwarfs in groups and clusters.
\end{abstract}




\section{Introduction}
\label{sec:intro}

Galaxies are seldom uniformly distributed within galaxy clusters. Groups and substructures have been detected observationally in cluster surveys using positions and velocities \citep{Gurzadvan_1998, Conselice_1998, Biviano_2002, Lisker_2018, Iodice_2019}, gravitational lensing \citep{Treu_2003,Natarajan_2009} and also in X-ray maps \citep{Owers_2009a, Owers_2009b, Zhang_2009, Jauzac_2016}. Given the typically large velocity dispersion associated with galaxy clusters, $\sigma \simeq 500$-$1000 \; \rm km \; \rm s^{-1}$, the probability of such associations to spontaneously occur after the galaxies are embedded in the gravitational potential of the cluster is negligible. Such associations of galaxies must, therefore, have already fallen in together as single units during the assembly of their cluster hosts.

This type of group infall arise naturally in hierarchical formation models such as the $\Lambda$CDM, where structure and substructure are self-similar. This means that not only the mass distribution within the host halo has a nearly universal profile regardless of halo mass \citep{Navarro_1996,Navarro_1997}, also the distribution of mass in substructures (i.e., satellites) that infall into those hosts can be described by a single function, provided that substructure masses are properly scaled by the host halo mass \citep{Giocoli_2008, Giocoli_2010, Yang_2011}. The latter is referred to as the ``unevolved satellite mass function" because it quantifies mass of substructure at infall and not at $z=0$, which may depend on the host \citep[see ][ for details]{vandenBosch_2005,Giocoli_2008}. This self-similar behavior in structure and infalling substructure follows in general for any model characterized by a scale-free power-spectrum of perturbation (like the $\Lambda$CDM) combined with the scale-free nature of gravity.

It is possible to quickly explore what this universal unevolved satellite mass function means for the assembly of different hosts. Taking Eq.2 in \citet{Giocoli_2008} we find that $\sim 2$ satellites are expected to infall with mass 10\% that of the host. Applied to the Milky Way (MW), this rightly predicts 1-2 subhaloes with infall mass $10^{11}$\msun\; comparable to that inferred for the LMC, the most massive satellite in the MW. This is in agreement with current estimates on the assembly history of the MW \citep{Busha_2011b, deLucia_2012MW, Boylan-Kolchin_2010}. If instead, one now applies this argument to a host halo like the Virgo cluster, assuming a virial mass $\sim 10^{14}$\msun, it predicts that 1-2 subhaloes with halo mass $\sim 10^{13}$\msun\; are expected to infall during the assembly history. This large halo mass, comparable to low-mass groups of galaxies, will be hosting more than one single galaxy and demonstrates that group infall is a clear prediction of $\Lambda$CDM. 

Considering lower mass subhaloes, the unevolved satellite mass function predicts $\sim 20$ objects with halo mass $\geq 1\%$ that of the host. For a cluster-mass host like Virgo, this means about $\sim 20$ MW-mass galaxies and above, with halo mass $\geq 10^{12}$\msun. This seems not too far off from observed stellar mass functions in Virgo for galaxies with $M_{\star} \geq 10^{10}$\msun\; \citep{Trentham_2002, Rines_2008} or from statistical studies in SDSS for groups in this mass range \citep{Yang_2009}. However, we have not accounted for tidal disruption and stripping, which will tend to lower the number of galaxies that survive at $z=0$ in this mass range. Therefore, some $\sim L_{\star}$ will be necessarily contributed from the larger mass groups with halo mass $\sim 10^{13}$\msun. 

Several studies have quantified the relevance of group accretion using simulations and semi-analytical models \citep{Berrier_2009, McGee_2009, Cohn_2012, deLucia_2012c, Roman_2017, Taylor_2004, Taylor_2005a, Taylor_2005b, Bahe_2019}.
For example, using N-body only cosmological zoom-in simulations of clusters \citet{Berrier_2009} measured about 25\% of MW-mass objects in clusters must be accreted as satellites instead of centrals from the field. This fraction is unknown for lower mass dwarfs. 

The connection between these theoretical predictions and observations is, however, not straight-forward. After these groups of galaxies enter the gravitational potential of the cluster, tidal forces will act to disrupt the coherence of the group \citep{Gonzalez-Casado_1994}, eventually mixing and erasing the typical signatures of common origin in the group members, such as spatial proximity and low mutual velocities. The ability to reconstruct, in present-day observations, the presence of galaxy groups that fell in together will depend on three factors: $i)$ the typical time-scale for disruption of the groups, $ii)$ the assembly history of the cluster --early formation means more time for tides to disrupt the groups-- and $iii)$ the multiplicity of the groups --having more galaxies allows for easier sampling of clustering in velocity or spatial scale. 

The third point mentioned above is automatically addressed by studying fainter galaxies, as low mass dwarfs dominate in numbers in any system. The situation for factors listed in $i)$ and $ii)$ is less clear. Previous work has indicated a large range of tidal disruption timescales, going from a very rapid dissociation of the groups \citep{Gonzalez-Casado_1994, Lisker_2018, Choque-Challapa_2019} to expecting several Gyr \citep{Vijayaraghavan_2015}. Partially, such discrepancies can be explained by the use of different definitions in the literature; for instance when the system is considered tidally disrupted or the exact definition of what is a group or association of galaxies at infall. However, some consensus exists: when substructure in position or velocity is {\it observationally} detected, it is interpreted as a young accretion event, typically suggesting less than $1$-$2$ Gyr since infall \citep{Lisker_2018}.

Here, our comparison to the MW can be used again to build up intuition. A scaled-down version of group infall is observed in our own galaxy with the LMC, which is expected to have brought along several fainter dwarfs that cluster in phase-space with the orbit of the LMC \citep{Donghia_2008,Sales_2011, Sales_2017, Jethwa_2016}. Observationally, the identification of dwarfs associated with the LMC is facilitated by two factors. First, detection of dwarfs in the MW goes down to the ultra-faint limit $M_{\star} \sim 10^3$-$10^4$\msun, allowing to sample several group members. Second, the Magellanic Clouds are on their first infall and have only recently fallen into the MW no more than 2 Gyr ago \citep{kallivayalil_2006, kallivayalil_2013}. 

In extragalactic systems that are megaparsecs away, projection effects in conjunction with shallower galaxy surveys make the identification of groups more difficult. Take for instance nearby clusters like Virgo or Fornax, where exquisitely deep surveys are able to map dwarfs down to $M_{\star} \sim 10^6$-$10^{7.5}$\msun \citep{Ferrarese_2020, Venhola_2017}. This is still several orders of magnitudes brighter than is possible in the MW.  Encouragingly, some evidence already exists, for example in the Virgo cluster, where samples of dwarfs with $r$-band magnitudes $-17 \geq M_r \geq -18$ show a skewed velocity distribution that has been attributed to a recent group infall event \citep{Lisker_2018}. Moreover, both systems, Virgo and Fornax have also obvious ongoing group infall in their outskirts, with the M49 group and Fornax A examples, respectively \citep{Binggeli_1987, Su_2019, Iodice_2017}. However, the detection of older accretion events deeper into the potential well of the clusters and/or groups with smaller masses remains an observational challenge, complicating the direct comparison to theoretical predictions.

Could group infall also play a role in explaining the different properties of cluster galaxies versus those in the field? Galaxies in clusters typically show lower star formation rates, older stellar populations and more spherically-dominated morphologies than field counterparts with similar mass. How much of this transformation is due to the infalling group environment versus the cluster environment is unclear, but there is exciting observational evidence in favor of ``pre-processing" in groups being an important factor lowering the star formation of some galaxies already before they join the cluster \citep{Roberts_2017, Sarron_2019, Mihos_2003, Vijayaraghavan_2013}. Bear in mind that only a fraction of the galaxies will infall as part of groups, and often further action by the cluster might be necessary to fully explain the low star activity of observed cluster galaxies.

On the other hand, group environments may be much more relevant on the {\it morphological transformation} of galaxies, by hosting mergers between group members. As was realized early on, the high velocity dispersion between galaxies in clusters make cluster environments unsuitable for mergers to occur \citep{Ostriker_1980, Fakhouri_2008,Fakhouri_2010a, Mihos_2003}. However, observational evidence of mergers in clusters do exist \citep[e.g. ][]{Moss_2006}. For instance, the presence of shell dwarf galaxies in the outskirts of Virgo \citep{Paudel_2017, Zhang_2020} is a vivid reminder that mergers can occur in galaxy clusters. 

Mergers have been postulated as a viable mechanism to transform morphology, in particular in galaxy types that are commonly found in galaxy clusters, such as lenticular (S0) galaxies \citep{Bekki_1998, Donofrio_2015, Eliche-Moral_2018} or blue compact dwarfs \citep{Bekki_2008, Zhang_2020}. Groups of galaxies infalling together will have shallower gravitational potentials than the cluster, allowing for lower velocity interactions to occur and, eventually, mergers  \citep[see e.g., ][]{Knebe_2006, Vijayaraghavan_2013, Bahe_2019}. Therefore, understanding the number of groups expected, their mass distribution and typical timescales for group disruption is deemed essential to understand the possible contribution of mergers to the formation of, for instance, S0 and blue compact dwarf galaxies while evaluating the need and contribution from additional secular and environmental formation channels for these morphologies. 

In this paper, we use the Illustris simulations to address the impact of group infall in the assembly of Virgo-like galaxy clusters. We introduce our sample and definitions in Sec.~\ref{sec:sims}. We characterize the fraction of galaxies in groups in Sec.~\ref{sec:acc}, the dynamical timescales for group disruption in Sec.~\ref{sec:evol} and the presence of mergers in Sec.~\ref{sec:mergers}. We summarize our main findings in Sec.~\ref{sec:concl}.

\begin{figure}
	\includegraphics[width=\columnwidth]{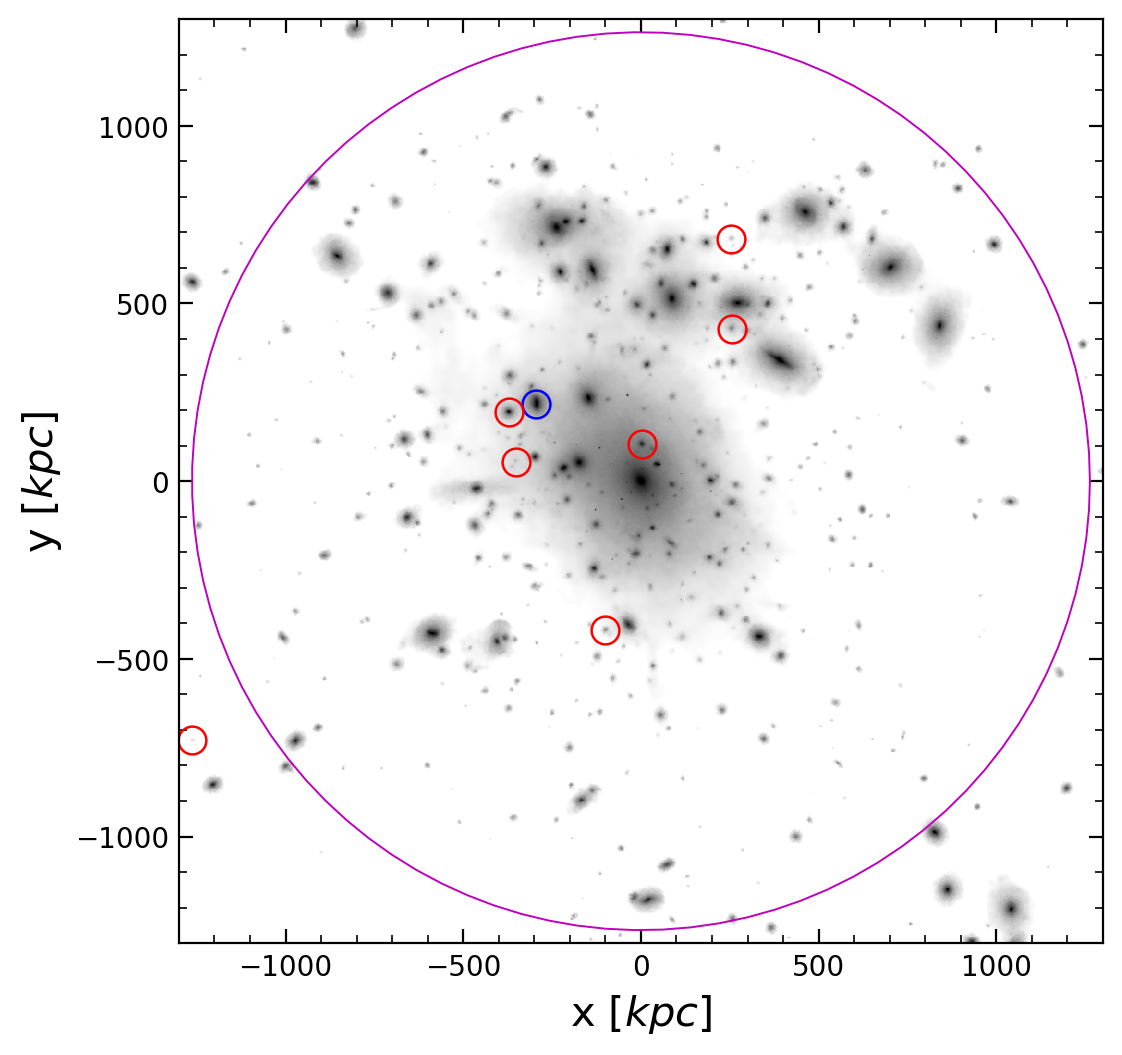}
    \caption{Projected stellar map at redshift $z=0$ of the most massive cluster in our Illustris sample (FoF0). This object has a virial mass $\rm{M_{200} = 2.32 \times 10^{14} \, M_{\odot}}$ and 232 satellite galaxies with $\rm{M_{\star} \geq 1.5 \times 10^8 \, M_{\odot}}$ inside the virial radius $r_{200} = 1.27$ Mpc, indicated by the magenta circle. We highlight with small colour symbols a set of galaxies that were part of a group before joining the cluster at $z = 1.67$. The blue circle shows to central of the group and the red circles its surviving satellites. Despite their common origin, little evidence of the past association for these galaxies remains at present-day. }
    \label{fig:cluster_z0}
\end{figure}


\section{Numerical Simulations}
\label{sec:sims}

We use the Illustris simulations \citep{Vogelsberger_2014a, Vogelsberger_2014b}, a set of cosmological hydrodynamical numerical simulations run with the {\sc arepo} code \citep{Springel_2010}. The Illustris suite consists of a $106$ Mpc on-a-side box simulated, in its highest resolution, with  $1820^3$ dark matter particles and an initially equal number of gas cells (Illustris-1). The corresponding mass per particle is $m_{dm}=6.3 \times 10^6 M_{\odot}$ for the dark matter and  $m_{gas}=1.3\times 10^6 M_{\odot}$ for the baryonic component. The gravitational softening is $\epsilon \sim 0.7$ kpc, although the hydrodynamics can reach a higher spatial resolution in the high density regions.

Illustris is evolved from initial conditions at redshift $z=127$ forwards in time until $z=0$ with cosmological parameters chosen to be consistent with results from Wilkinson Microwave Anisotropy Probe WMAP9 \citep[]{Hinshaw_2013}. The main astrophysical processes shaping galaxy formation are included, such as the effects of cooling and heating of the gas, star formation, stellar and metallicity evolution and  feedback from stars as well as black hole sources. Detailed descriptions of the model can be found in previous papers \citet{Vogelsberger_2013, Vogelsberger_2014b, Genel_2014, Sijacki_2015, Nelson_2015}.

Structures are identified in a two-step process run on-the-fly with the simulation. First, groups are identified using the Friends-of-Friends algorithm, FoF \citep{Davis_1985} based solely on the spatial distribution of particles. Subsequently, {\sc subfind} is run using 6D information to identify galaxies and subhaloes that are self-bound structures within those groups \citep{Springel_2001, Dolag_2009}. Galaxies at the center of the gravitational potential in each group are flagged as ``centrals", whereas all the remaining substructures identified in each group are flagged as ``satellites". We use the {\sc sublink} trees to follow the temporal evolution of identified galaxies \citep{Rodriguez-Gomez_2015}.

In this paper, we select the $10$ most massive haloes in the Illustris-1 box at $z=0$, corresponding to galaxy clusters with virial mass $M_{200} \sim 10^{14}$\msun\; (exact range $14.02< \rm log(M_{200}/M_\odot)<14.37$). We define virial quantities based on an overdensity contrast equal to $200$ times the critical density of the universe. The virial radius of our clusters are in the $0.97< \rm log(r_{200}/Mpc)<1.26$ range with typical velocity dispersion within those radii $899<\sigma_{200}/\rm km \, \rm s^{-1}<1073$. Within these $10$ host haloes, we follow the evolution of all galaxies with stellar mass $M_{\star} \geq 1.5 \times 10^8$\msun, corresponding to an average of $\sim 120$ stellar particles in the lowest mass objects. Galaxy quantities such as stellar mass, $M_{\star}$, and gas mass, $M_{\rm gas}$, are measured within twice the half mass radius of the stars (using {\tt SubhaloMassInRadType} in the {\sc subfind} catalogs), while dark matter mass is defined as all dark matter particles gravitationally bound to a given subhalo (using the field {\tt SubhaloMassType}).

\begin{figure*}
	\includegraphics[width=\columnwidth]{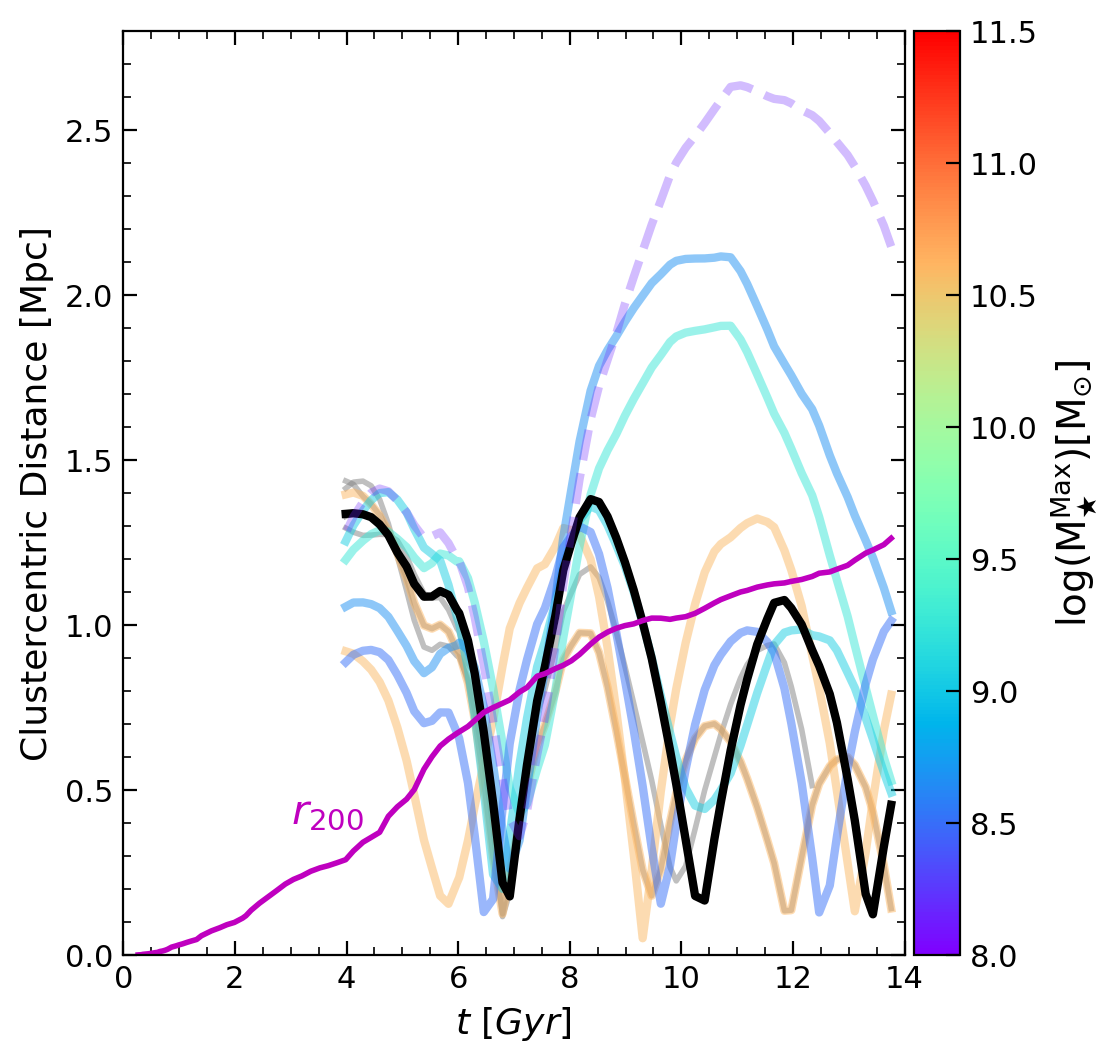}
	\includegraphics[width=\columnwidth]{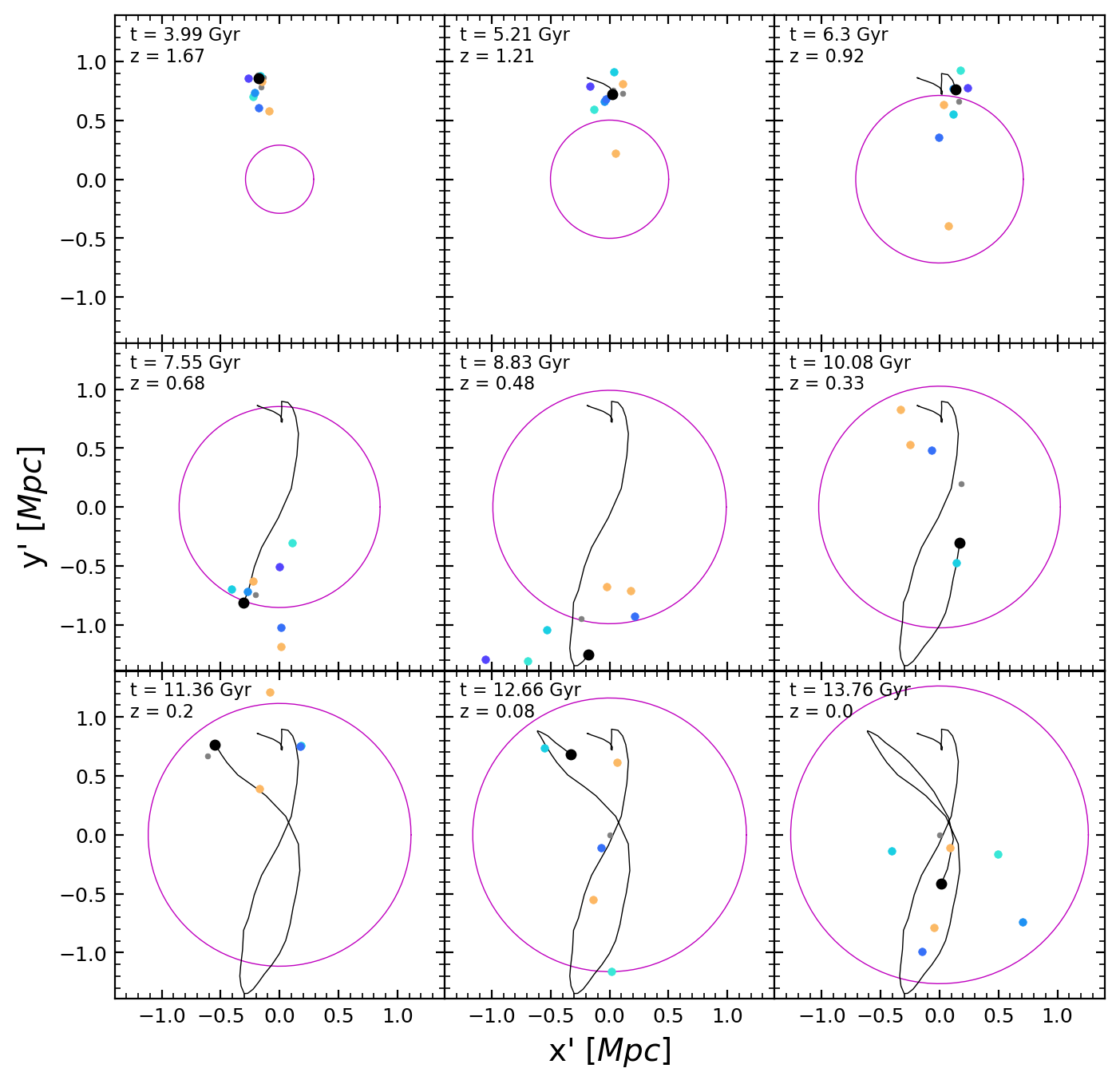}
    \caption{{\it Left:} orbit versus time for the infalling group of galaxies highlighted in Fig.~\ref{fig:cluster_z0} after their infall into FoF0. The central galaxy in the group is shown in black, and the satellites are colour-coded according to their maximum stellar mass (see bar on the right). Note the propelling of three ``escapees" after the first pericenter passage, that takes them well beyond the virial radius of the host cluster (magenta curve). At $z=0$ there is little coherence left among these orbits. {\it Right:} XY projections of the group at different times (see legends) with the trajectory of the central galaxy outlined with the thin black curve. The system has been rotated according to the angular momentum of the orbit of the central galaxy (pointing in the z-direction) and the colours are the same than in left panel. Although the group starts very spatially clustered, the signal is appreciably weakened after the first and second pericenter within the cluster.}
    \label{fig:orbits}
\end{figure*}

\subsection{Galaxy Clusters and Group Infall}
\label{sec:sample}

We show in Fig.~\ref{fig:cluster_z0} a stellar projection of our most massive galaxy cluster (FoF0), at present day, where the magenta large circle indicates the virial radius for this object. A total of 232 galaxies with $M_{\star} \geq 1.5 \times 10^8$\msun\; are identified within $r_{200}$ in FoF0 at $z=0$, which are visible here as substructure in the gray-scale map. The {\sc sublink} merger trees allow us to trace backwards on time the evolution of each galaxy until their time of infall, $t_{\rm inf}$, defined here as the previous snapshot when they join the same FoF group of the cluster progenitor. This definition typically places infalling galaxies at a distance of $2$-$3$ times the virial radius of the cluster. It has been shown that environmental effects such as halo stripping start well beyond the virial radius \citep[e.g. ][]{Behroozi_2014} and peak at around $\sim 2 r_{200}$, providing support for our $t_{\rm inf}$ definition.  

We consider infalling ``galaxy groups'' as any FoF group with at least $2$ (but up to $80$) galaxy members above our resolution cut on $M_{\star}$. Galaxies are classified as ``centrals" if they were the central object of their own FoF group at $t_{\rm inf}$, or as ``satellites" otherwise. With this definition, a galaxy will be considered accreted as a satellite even if they were outside the virial radius of their infalling group (although we briefly discuss the consequences of assuming a more strict criteria in Sec.~\ref{sec:acc}). 

From this point of view, our results characterize well the infall of loose associations of galaxies rather than fully virialized structures. After infall, tidal forces in the cluster will tend to dissolve these associations over time. For example, Fig.~\ref{fig:cluster_z0} highlights with red and blue circles all galaxies of a once single-group, whose members lay today at $z=0$ quite mixed within the cluster. Red indicate those galaxies accreted as satellites of this group, while blue corresponds to the central of the group. The virial mass of this substructure at infall is $M_{200}(t_{\rm inf})=M_{200}^{\rm inf} = 4.69 \times 10^{12} $\msun\; and the stellar masses for the associated galaxies at infall are in the $\rm log(M_{\star}/$\msun$)=[8.33-11]$ range. 

The orbital evolution with time of these group members are shown in detail in Fig.~\ref{fig:orbits}. The left panel shows the distance versus time for each galaxy in the group with respect to the cluster center and emphasizes an initial coherence during the infall which is later weakened over time. This is particularly true after the first pericenter passage around $t\sim 7$ Gyr. Interestingly, some group members may gain energy due to several-body interactions, resulting in the display of odd orbits that place them today outside the virial radius of the cluster (dashed curve in Fig.~\ref{fig:orbits}, bottom left red circle in Fig.~\ref{fig:cluster_z0}). These unorthodox orbits have been found in the literature to be common in simulations of group infall {\citep[e.g., ][]{Balogh_2000, Sales_2007b, Ludlow_2009} and related to the ``backsplash" radius region in observations and simulations \citep{Pimbblet_2011,Muriel_2014, Diemer_2017}.

The right panel in Fig.~\ref{fig:orbits} shows an XY projection of these orbits, and illustrates the disruption of the group over time as the satellite companions deviate from the trajectory of the central in the group (thin black line). The combination of different assembly histories for each of the clusters combined with the timescales over which these tidal disruptions of galaxy groups occur will determine the level of substructure --in both, position and velocities-- that is expected in clusters within $\Lambda$CDM. In what follows, we study statistically the infall of galaxy groups onto our $10$ simulated clusters and characterize their role in the assembly and dynamical evolution of their cluster hosts.

\begin{figure}
	\includegraphics[width=\columnwidth]{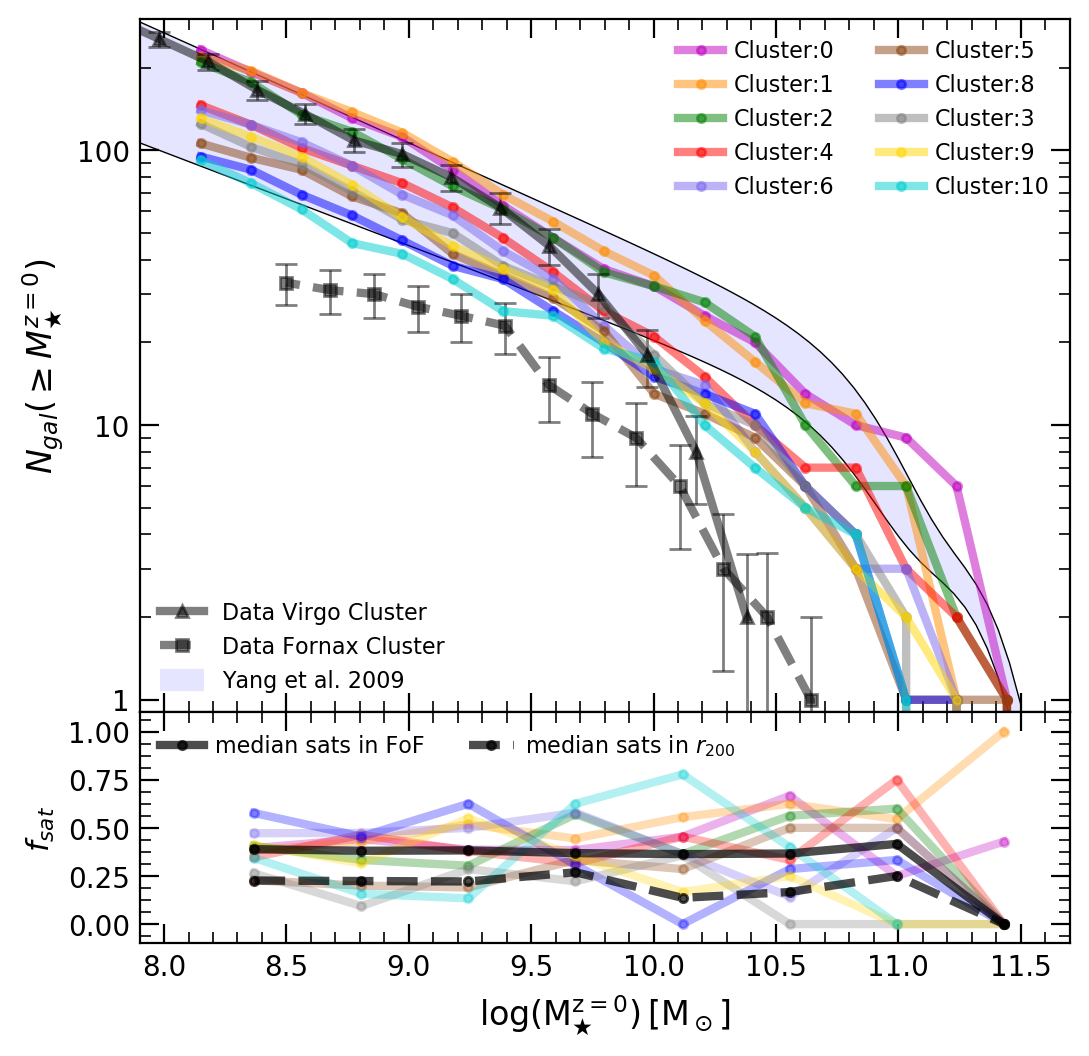}
	\caption{Cumulative galaxy stellar mass function at $z = 0$ for our $10$ simulated clusters (colour lines). The predicted number of satellites at $z=0$ in Illustris is in good agreement with observational data in similar mass ranges: the Virgo cluster \citep[within $\sim 1 \, Mpc$  from M87 in solid gray, ][]{Rines_2008}, Fornax \citep[inside $\sim 0.7 \, Mpc$ in dashed gray, ][]{Sarzi_2018} and in shaded blue SDSS groups and clusters \citep[range $ \log(M_h/M_{\odot}) \in (13.8,14.4)$ in ][]{Yang_2009}. The lower panel quantifies the fraction of these galaxies in each $M_{\star}$ bin that entered the clusters as satellites of groups with $N_{\rm gal} \geq 2$ members. We find a median of $38\%$ entered as satellites with little dependence on stellar mass (solid black curve), although variations from cluster-to-cluster are large (coloured lines). The dashed black line shows the median $f_{\rm sat}$ in the case of a more strict definition of satellite infall (see text for more details).}
   \label{fig:mass_func}
\end{figure}

\begin{figure}
	\includegraphics[width=\columnwidth]{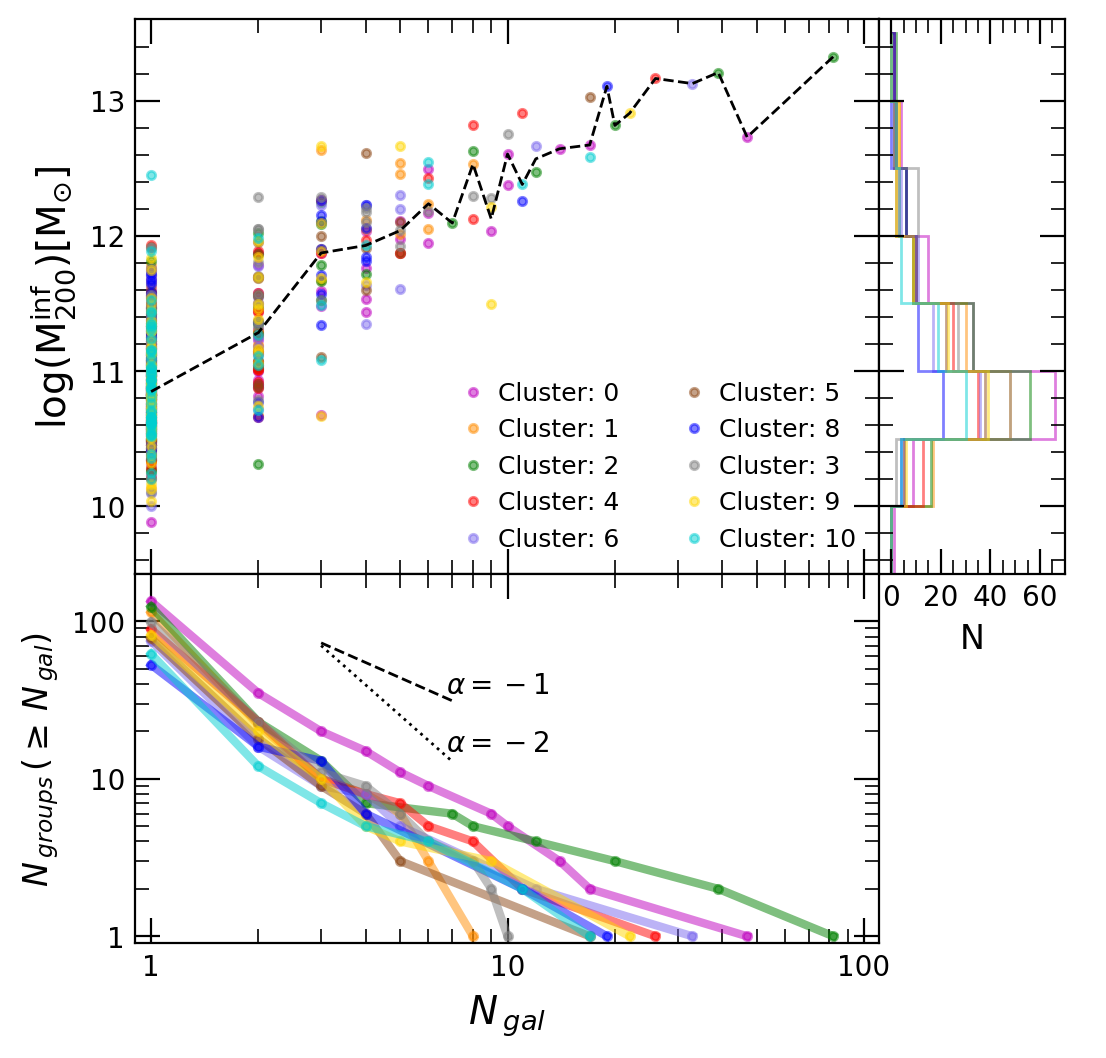}
	\caption{Multiplicity of infalling galaxy groups as a function of their virial mass. Multiplicity is defined as the number of galaxies, $N_{\rm gal}$, above our cutoff mass  $\rm{ M_{\star} \geq  1.5 \times 10^8 \,  M_{\odot}}$ that enter as part of the same group. Low mass groups will typically bring only one galaxy, but groups with masses $M_{200}^{\rm inf} \geq 2 \times 10^{12}$\msun\; may contribute $\sim$10 to 80 galaxies. The bottom panel shows the cumulative multiplicity function (how many groups with $N_{\rm gal}$ larger than $x$) for each of our clusters. The histogram on the right shows how many objects of a given $M_{200}^{\rm inf}$ are found per cluster.} 
   \label{fig:ngal}
\end{figure}

\section{Accretion of galaxies in groups}
\label{sec:acc}

We start by validating the stellar mass function of our Illustris simulated galaxy clusters with observational constraints of Virgo and Fornax clusters (estimated virial masses: $\rm M_{200} = 7.0 \pm 0.4 \times 10^{14} \, M_{\odot}$ and $\rm M_{200} = 7 \pm 2  \times 10^{13} \, M_{\odot}$, respectively \citep{Karachentsev_2014, Drinkwater_2001}). Our clusters contain between $92$ and $232$ galaxies with $M_{\star} \geq 1.5 \times 10^8$\msun\; that are today within $r_{200}$. Fig.~\ref{fig:mass_func} shows the (cumulative) stellar mass function of galaxies in each simulated cluster (solid colour lines). The scatter mostly corresponds to the range of virial masses selected and the different assembly histories sampled in the simulation. The numbers and distribution of $M_{\star}$ in our objects seem bracketed by measurements in Virgo \citep[black triangle symbols, ][]{Rines_2008} and Fornax \citep[black square symbols, ][]{Sarzi_2018}}. Simulations also agree well with estimates from groups and clusters in SDSS \citep[][]{Yang_2009} using the set of fitting parameters corresponding to the $\rm log(M_{200}/M_{\odot})=[13.85-14.39]$ range that best approximates our systems. 

The merger trees allow us to measure the fraction of the simulated galaxies that fell into their clusters as satellites of a galaxy group. We show this in the bottom panel of Fig.~\ref{fig:mass_func}, where the colours correspond to the same individual clusters as the main panel and the black thicker solid line indicates the median of all clusters combined. 

A significant, although sub-dominant, fraction of galaxies enter the cluster as part of larger associations, with appreciable cluster-to-cluster scatter. Combining all our systems, we find that $38 \pm 15 \%$  of galaxies enter as satellites, with no clear dependence on galaxy stellar mass up to $M_{\star} \sim 10^{11}$\msun, after which the satellite fraction plummets to near zero. Dwarf galaxies with $M_{\star} \sim 10^8$\msun\; are equally likely on average to be a satellite at infall as more massive galaxies in the $\sim L_{\star}$ range.

The exact values quoted above depend on the definition of what is considered a ``group" at infall. As mentioned in Sec.~\ref{sec:sims}, we define as satellites all objects that are part of a FoF group. Note that other more strict definitions have been used in the literature before, for example, by requiring that satellites are within the virial radius of a bound group \citep[e.g., ][]{Berrier_2009, Choque-Challapa_2019}. For comparison, we find that when such criteria is used, our satellite fractions are about a factor of two smaller  ($22 \pm 12 \%$) than with our default definition, see dashed line in bottom panel of Fig.~\ref{fig:mass_func}. This is in excellent agreement with results reported in \citep{Berrier_2009} for their $L_{\star}$ galaxies. 

The break down of the accreted groups in mass and galaxy multiplicity is shown in Fig.~\ref{fig:ngal}. We use the virial mass at infall $M_{200}^{\rm inf}$ to quantify the masses of the FoF groups, which show, as expected, a clear correlation with the number of members in the group. In general, singletons and pairs are by far the more common accretion events \citep{Choque-Challapa_2019} for galaxies with $M_{\star} \geq 1.5 \times 10^8$\msun, but higher multiplicity is also common: about $\sim 10$ groups with $N_{\rm gal} \geq 3$  galaxies are predicted during the assembly of Virgo-like clusters. (The quoted numbers correspond to the requirement that at least 1 member of the group will survive at $z=0$.) 

The variations from system to system are larger on the more numerous groups, where we expect a few (albeit likely) events brining $\sim 10$ galaxies, with some extreme cases contributing up to $80$ members (green curve, cluster 2). This scatter reflects the expected differences on the particular assembly history of each of the host clusters. The median virial mass for $N_{\rm gal} = 2$ accretions is $1.9 \times 10^{11}$\msun\; suggesting that associations of dwarfs being accreted into clusters may be rather common. However note the significant vertical scatter at $N_{\rm gal} = 2$, indicating that also MW-mass groups might bring-in only one companion (besides the central) as a result of the large halo to halo variations. 


\section{Dynamical evolution of accreted groups}
\label{sec:evol}

\begin{figure}
	\includegraphics[width=\columnwidth]{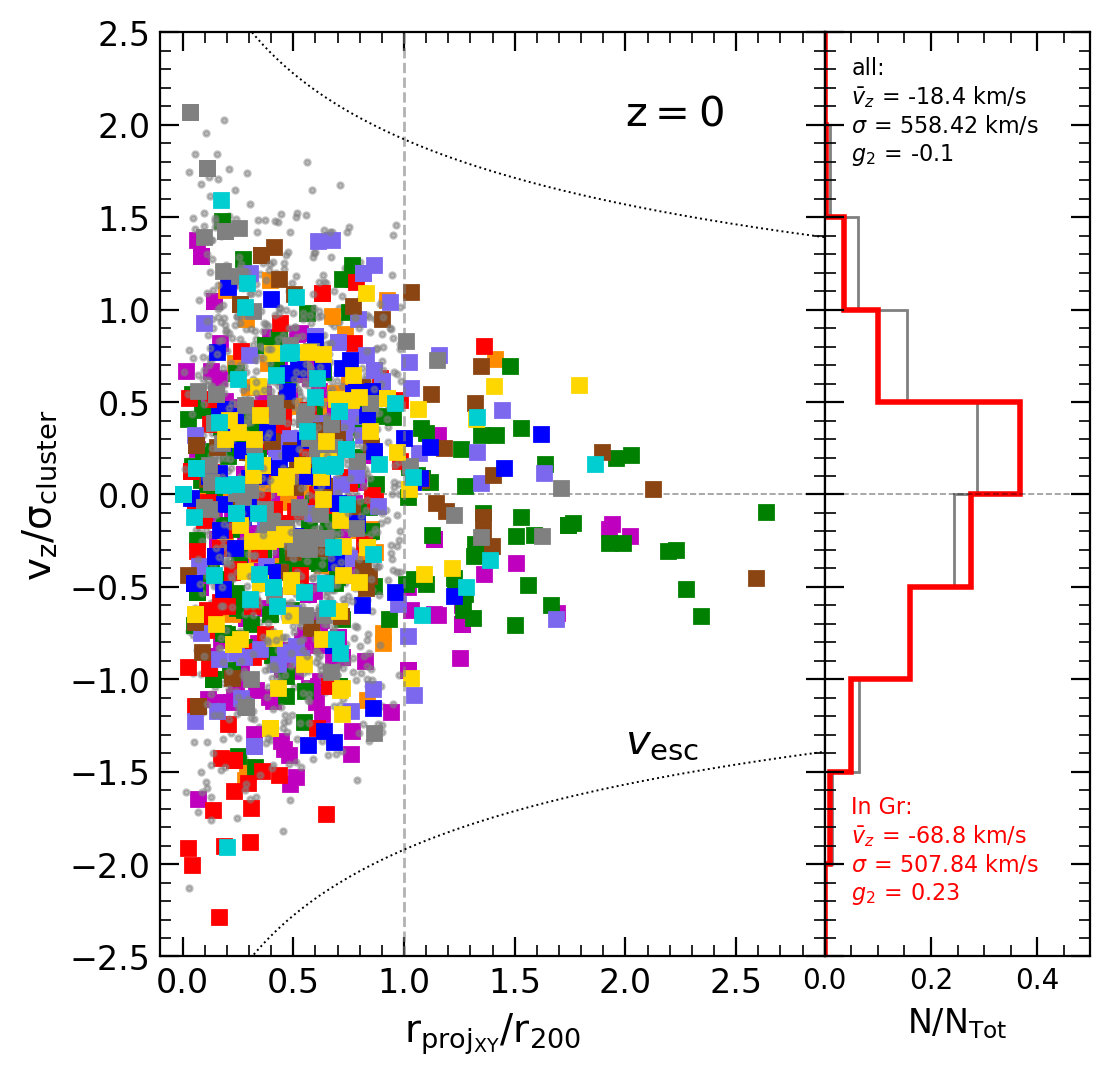}
    \caption{Projected distance vs. line-of-sight velocity for galaxies in clusters at $\rm{z = 0}$ (gray) with colours highlighting objects accreted as part of groups with  $\rm{N_{gal} \geq 2 }$ (colour code is the same as in Fig.~\ref{fig:mass_func}). All galaxies are confined within the average escape velocity of the clusters (dotted curve). The vertical histograms on the right show that galaxies that infall in groups have a similar velocity distribution than normal galaxies when taken as a whole. Quoted values correspond to the mean, dispersion and skewness of each distribution. Signatures of group infall, if present, dissolve quickly within the clusters.}
    \label{fig:phase_space}
\end{figure}

Despite the prevalence of group accretion shown in Sec.~\ref{sec:acc}, evidence of substructure within clusters at $z=0$ is not abundant, at least in phase space. Fig.~\ref{fig:phase_space} shows for our simulated clusters the velocity vs. clustercentric distance of galaxies at present day, projected along a random line of sight to facilitate comparison to observations. Gray dots indicate galaxies that have fallen in as single objects, while colours indicate those aggregated as part of groups with $N_{\rm gal} \geq 2$ members. Dotted line shows the expected escape velocity assuming an NFW profile \citep{Navarro_1996} and concentration {$c=5.24$ following \citet{Duffy_2008} for a $M_{200}=1.6 \times 10^{14}$\msun\; halo. The histograms shown in the right subpanel suggest that the global distribution of these two populations is not sufficiently different at $z=0$ to distinguish substructure infall in this space. We have explicitly checked that this holds true also in a cluster-by-cluster basis.

Observationally, asymmetries in projected phase-space coordinates have been attributed to the accretion of substructure. For example the Virgo cluster, where dwarf galaxies with r-band absolute magnitude $-17 \geq M_r \geq -18$} show hints of an unrelaxed state, has been proposed as the smoking gun evidence for substructure infall \citep[e.g., ][]{Lisker_2018}. Such strong signature is not present in our simulations despite the important role of group infall in our systems. Here we have explicitly checked that the velocity distributions remain similar even when taking only low mass galaxies ($M_\star \leq 10^9$\msun\; assuming $M_r=-17$ and mass-to-light ratio $\sim 1$), as done in Virgo. One way to reconcile our theoretical expectations with the detection of substructure in observations is to assume that the clustering, either in position or velocities, of the infalling galaxy groups is rather quickly dissipated. 

Qualitatively, a hint of this behavior can be seen from the positions of galaxies in our example infalling group in Fig.~\ref{fig:orbits}, where the coherence in the orbital structure of the group is lost after the first pericenter at $t \sim 7$ Gyr. As a consequence, the distance between the galaxy members increases substantially, losing the clustering signature expected for substructures. This group dispersal effect even includes galaxies going back outside of the virial radius of the cluster at $z=0$ while the rest of the group distributes within $r_{200}$. Naturally, a weakening of the correlation in velocities between galaxy members is also expected. 

\begin{figure}
	\includegraphics[width=\columnwidth]{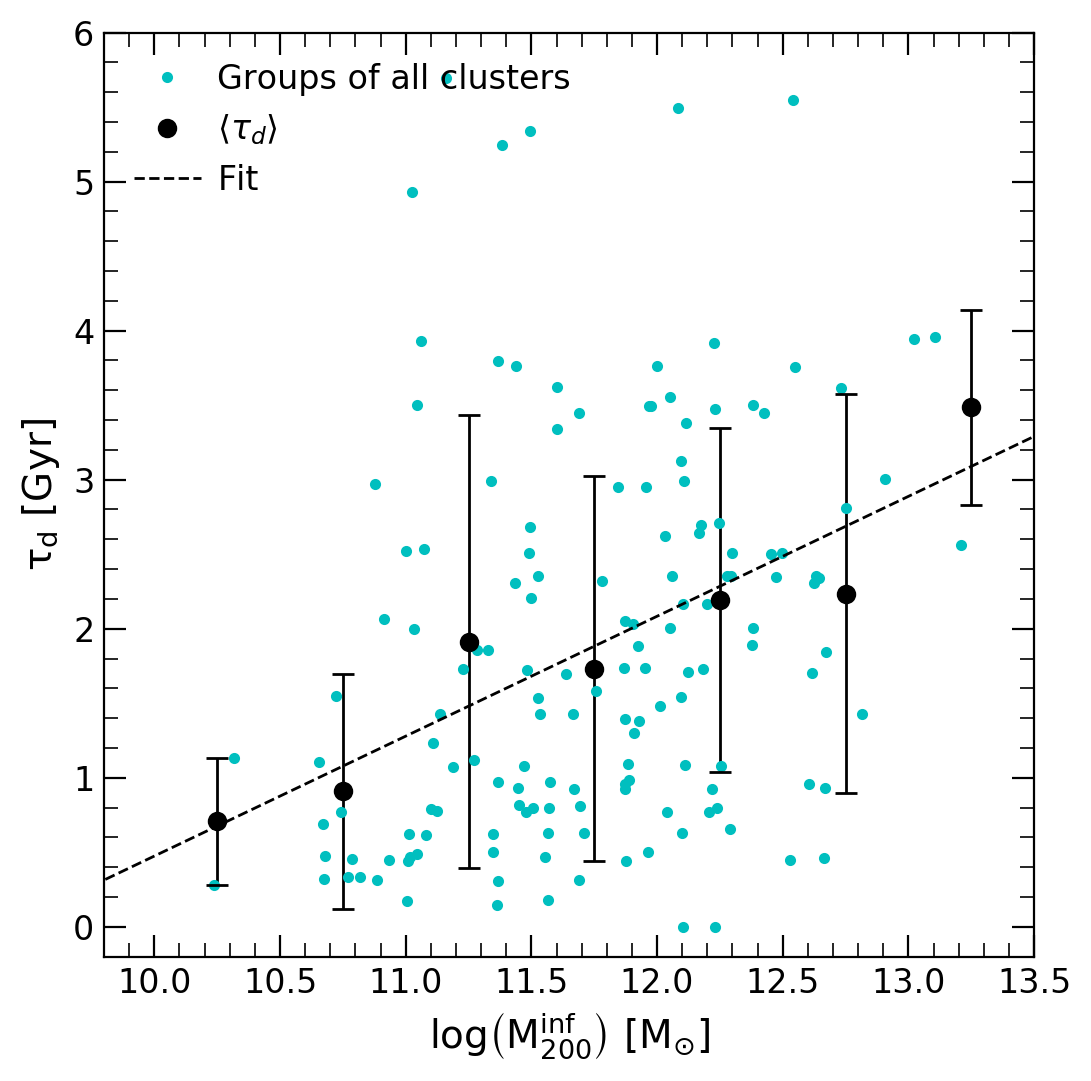}
	\caption{Time $\tau_d$ that is needed for each group to double their radial size, $\sigma_r$, as a function of the infall mass of the group. Individual groups identified in the 10 simulated clusters are shown in cyan circles with the median $\tau_d$ in $M_{200}^{\rm inf}$ bins shown in black symbols and bars for the standard deviation. There is a very weak dependence with the mass of the form: $\tau_d = \rm{ \alpha \, log(M^{inf}_{200}/M_\odot) + \beta}$, with $\alpha = 0.8 \pm 0.1$ and $\beta = -7.6 \pm 1.5$ (dashed line), but the trend is greatly superseded by the large scatter at each group mass. The large dispersion is attributed to the different orbits, infall times and number of members in each group. 
}
    \label{fig:time}
\end{figure}

The stretching in space due to the tidal disruption of groups by the clusters can be measured by tracking the evolution of a characteristic size for each galaxy association after infall. We define $\sigma_r$ as the r.m.s dispersion of distances of all galaxies in each group, where $\sigma_r = \sqrt{\sigma_x^2 + \sigma_y^2 + \sigma_z^2}$, with $\sigma_x = \sqrt{\langle x^2_i \rangle - \langle x_i \rangle^2}$ and the same for $\sigma_y$ and $\sigma_z$. More specifically, to quantify the timescales involved in group disruption, we measure the time $\tau_d$ that it takes for a given group to double its size $\sigma_r$ compared to that they had at infall time. We have explicitly checked that using other metrics to quantify size-evolution give quantitative similar results as using $\sigma_r$ (see Fig.~\ref{fig:appendix_disruption} in Appendix for a discussion). 

We show in Fig.~\ref{fig:time} this ``disruption" timescale $\tau_d$ as a function of the virial infall mass of our groups. Two points are worth highlighting. First, the size transformation occurs over short timescales, $\tau_d \sim 1$-$3$ Gyr, albeit with significant scatter. This means that, if substructure is identified in observations of galaxies within clusters, it is likely associated to a relatively recent accretion event \citep[see also, ][ for similar conclusions]{Choque-Challapa_2019}. However, the scatter also allows for less common cases where the spatial clustering may last over longer time periods ($\sim 4$-$5$ Gyr). 

Second, we find a weak trend with mass: low mass groups will double their spatial extend in about $\sim 1$ Gyr after infall, while more massive groups take on average $\tau_d \sim 3$ Gyr. These results can be interpreted as low mass groups being less resilient to the tides from the central cluster, while the self-gravity of more massive groups allows them to remain bound for longer times after infall. The median trend is well fit by a relation: $\tau_d = \rm{ \alpha \, log(M^{inf}_{200}/M_\odot) + \beta}$, with $\alpha=0.8 \pm 0.1$ and $\beta = -7.6 \pm 1.5$, where uncertainties correspond to standard regression errors. However, there is a large scatter in this relation that might reflect the different accretion times and orbits of groups with similar $\rm{M^{inf}_{200}}$. 

\begin{figure}
    \includegraphics[width=\columnwidth]{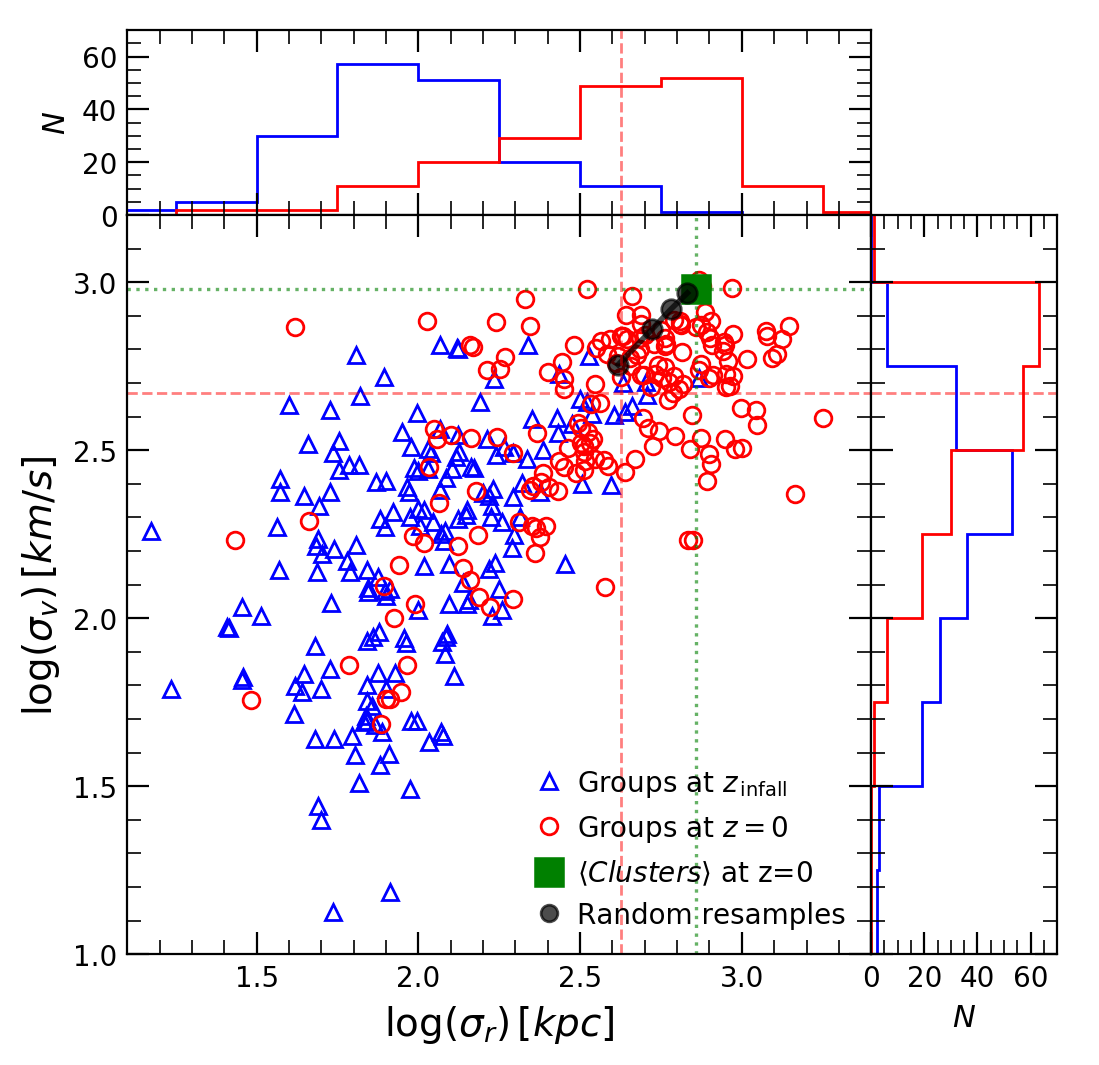}
	\caption{Time evolution of groups in characteristic size, $\sigma_r$, vs. their typical velocity dispersion, $\sigma_v$. Values at infall for each group are shown in blue while $z=0$ are shown in red. Groups after infall tend to thermalize with the state of the cluster, systematically increasing their sizes and velocities (see blue vs. red histograms along each axis). However, despite of the dynamical heating, groups tend to remain colder and more compact than the host cluster itself (compare the median of groups in dashed red lines with the median virial cluster values shown by thin dotted green lines and filled green square). The filled black circles symbols connected by a continuous line show the average $\sigma_r$ and $\sigma_v$ of 100 realizations with N = 2, 3, 5 and 20 randomly selected galaxies in the simulated clusters that are {\it not} accreted as part of groups. The distribution of red symbols (group infallers) is clearly shifted towards lower $\sigma_r$ and $\sigma_v$ than these random samples, confirming that group infall provides a special kinematical environment for members compared to the rest of the relaxed cluster population.}
    \label{fig:sigmas}
\end{figure}

Besides the spatial distribution of the group, the internal velocities of galaxy members are also changed by the dynamical evolution within the cluster. Following our definition of $\sigma_r$, we define $\sigma_v$ as the velocity dispersion between galaxies in the same group. Fig.~\ref{fig:sigmas} shows $\sigma_r$ vs. $\sigma_v$ for groups at infall (blue triangles) compared to the same groups at $z=0$ (red circles).
There is a clear evolution in both, size and velocity after infall. The median size of groups increases from $\rm{log(\sigma_r)} \sim 1.99$ at infall to $\sim 2.63$ at $z =0$, corresponding to $\sim \times 4.4$ average increase in size at present day. The velocity dispersion between galaxy members also increases by a factor $\sim \times 2.5$ in the same time period (with medians $\rm{log(\sigma_v)} \sim 2.27$ at infall compared to $\sim 2.67$ at present day). Groups get less spatially clustered and increase their velocity dispersion with time.

How large and how hot these groups become at $z=0$? Naively, one would expect that the size of the cluster and its velocity dispersion are natural boundaries to these quantities. We show in Fig.~\ref{fig:sigmas} (green square) the average size (virial radius) and the average velocity dispersion of our clusters. Additionally, for better reference, we also show the median $\sigma_r$ and $\sigma_v$ that one would measure by taking random groups of  N=2, 3, 5 and 20 galaxies in our clusters that did not infall as part of a group (filled black circles are the median of $100$ random samplings). By construction, these filled circles represent the expected closest-to-virialized $\sigma_r$ and $\sigma_v$ in these clusters and demonstrate that $\gtrsim 5$ members should be expected to trace average sizes and velocity dispersions in the cluster if properly relaxed. 

Instead, we find that galaxies that fell in as part of groups are today at $z=0$ typically below the velocity of these random samples (red symbols), suggesting that although the dynamical evolution in the cluster tend to erase the dynamical identity of substructures, groups that have fell in remain kinematically colder than the surrounding cluster. The same is not exactly true for the spatial distribution of groups, which may even exceed the virial radius of the cluster (red points to the right of the dotted green vertical line). This is the case for some groups where galaxy members were ejected outside $r_{200}$, as the group showcased in Fig.~\ref{fig:orbits}. The median $\sigma_r$ of galaxies in groups is still more clustered that the random samples (see vertical dashed red line), but the effect in position is less systematic than in velocity.

\section{Mergers of galaxies in groups}
\label{sec:mergers}

The lower velocities between galaxies associated with the infalling groups found in Fig.~\ref{fig:sigmas} may play a vital role in facilitating the conditions for mergers to occur within (or in the outskirts) of massive clusters. This is perhaps important to help explain the observational evidence for young mergers in nearby galaxy clusters like Virgo \citep[e.g., ][]{Zhang_2020}. We follow the assembly of our $10$ clusters in Illustris and identify possible merger events, their mass ratios, times and location with respect to the cluster center. 

\begin{figure}
	\includegraphics[width=\columnwidth]{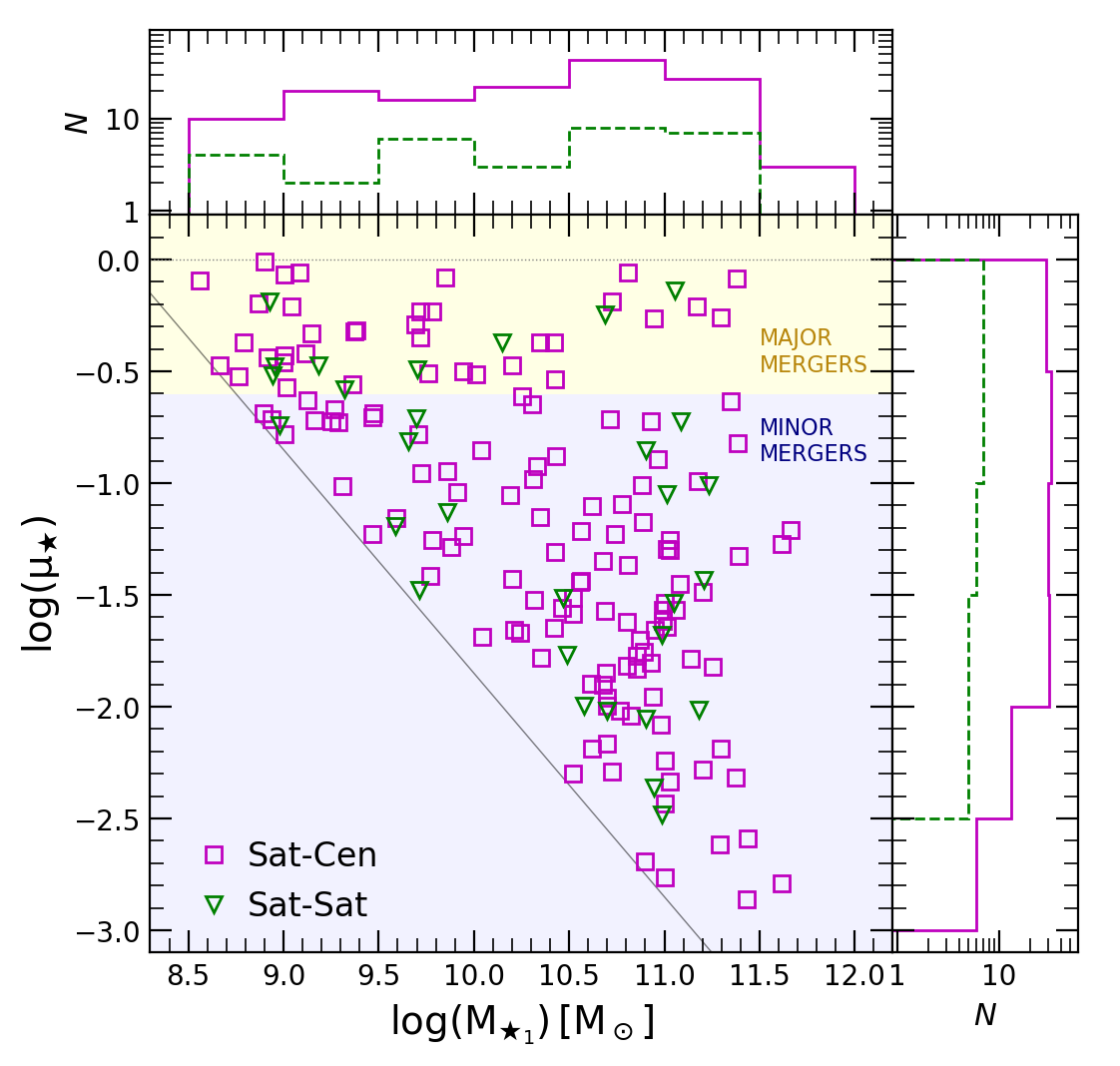}
	\caption{Stellar mass ratio $\mu_{\star}$ of the $171$ identified mergers in and around clusters as a function of the stellar mass of the most massive companion pre-merger $M_{\star_1}$. All mergers found are hosted by groups that fell into the cluster, either between a satellite galaxy and the central of the group (Sat-Cen, magenta squares) or between two satellites in the group (Sat-Sat, green triangles). We adopt the mass ratio $\mu_{\star} = 0.25$ to distinguish between major (yellow) and minor (purple) mergers following \citet{Rodriguez-Gomez_2015}. A significant fraction ($26.3\%$) of the mergers occurring in clusters are major according to this definition.}
    \label{fig:mergers_mass}
\end{figure}

We find a large number of galaxies that merge to the central galaxy in the cluster, referred to as ``brightest cluster galaxy", or BCG. These mergers build the stellar mass of the BCG as well as the intracluster light component and are not the main focus of this paper \citep[see for instance ][ for a quantification of accreted component in BCGs]{Rodriguez-Gomez_2015}. Instead,  we identify mergers that occur near or within the clusters, and that involve only satellite galaxies in the clusters. We find an average of   $\sim 17 \pm 9$ mergers per cluster above our resolution limits, where the quoted uncertainty corresponds to the r.m.s of our clusters sample. 

This confirms that indeed mergers can occur in clusters environments despite the high velocity of galaxies in such environment. Most importantly, {\it all mergers identified occur in galaxies that fell into the clusters as part of groups with $N \geq 2$ members}. In other words, mergers in clusters do not occur among galaxies that infall as singletons, in agreement with expectations based on the typical high velocity in clusters \citep{Ostriker_1980}.

Fig.~\ref{fig:mergers_mass} shows the stellar mass ratios $\mu_{\star} = M_{\star_2}/M_{\star_1}$ of all mergers detected, where the stellar mass of the most massive galaxy involved is defined as $M_{\star_1}$ and the secondary is $M_{\star_2}$. We follow the convention introduced in \citet[][]{Rodriguez-Gomez_2015} and record both masses at the time when the secondary has its maximum mass, to avoid artificial lowering of the mass ratio due to subsequent stripping before the merger \citep[see ][ for a detailed discussion]{Rodriguez-Gomez_2015}. 

We further divide the mergers in Fig.~\ref{fig:mergers_mass} into two kinds:  those between a satellite and the central of the group (Sat-Cen, magenta squares), and those involving two satellites within a group where neither is the central (Sat-Sat, green triangles). The distributions do not differ substantially, except that mergers with the centrals of the groups are more likely ($82\%$). 

Following \citet{Rodriguez-Gomez_2015} we consider galaxy mergers with stellar mass ratio $\mu_\star \gtrsim 0.25$ as major mergers. As expected from numerical limitations, we can only follow major mergers in our dwarf galaxies regime ($M_{\star} \leq 10^9$\msun) while for $M_{\star} \geq 10^{10.5}$\msun\; we resolve well into the minor mergers events. Sat-Sat and Sat-Cen show similar distributions for major and minor mergers (see vertical histogram on the right of Fig.~\ref{fig:mergers_mass}).

\begin{figure}
	\includegraphics[width=\columnwidth]{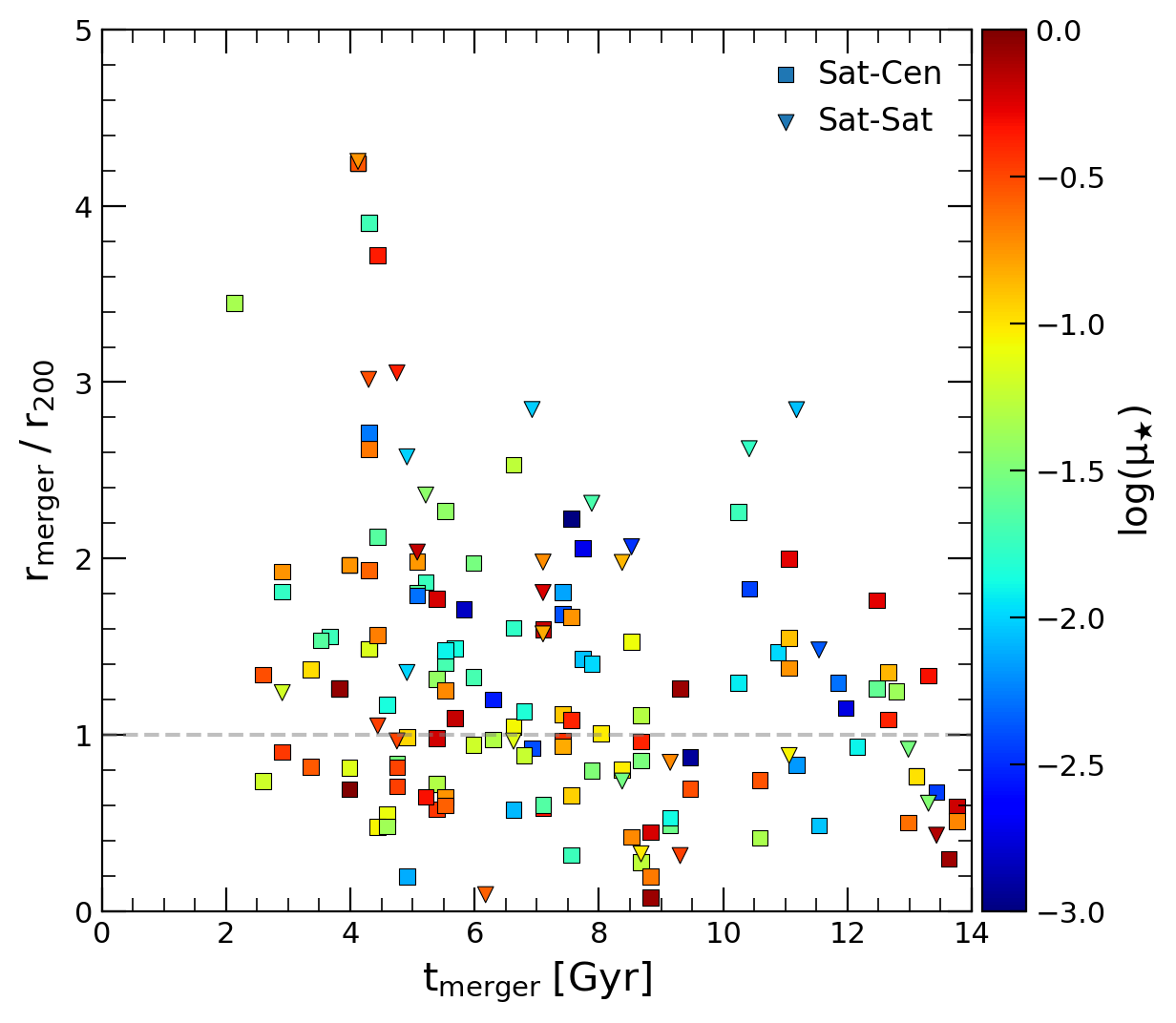}
    \caption{Distance (from the center of the clusters) versus time where the mergers occur. Symbols are colour-coded by stellar mass ratio $\mu_{\star}$ (see colour bar). The gray horizontal line indicates $r = r_{200}$. Around $\sim 40 \%$ of the mergers occur inside the virial radii of the clusters with no significant correlation with time or mass ratio.} 
    \label{fig:mergers_dist}
\end{figure}

Mergers may occur anywhere from the time of infall until today and they can therefore happen also in the immediately surrounding region of the clusters ($1 \leq r/r_{200} \leq 2$-3). We study the distribution of time and location of these mergers in Fig.~\ref{fig:mergers_dist}, where symbols correspond to each of the Sat-Sat and Sat-Cen merger identified and colours indicate mass ratio $\mu_{\star} = M_{\star_2}/M_{\star_1}$ using the colour coding shown in the vertical bar. Although the majority of these mergers happen within the group environment before crossing the virial radius, still a significant fraction of mergers, $\sim 40\%$, happen within $r_{200}$ (corresponding to an average number per cluster of $7 \pm 3$ throughout its evolution). Moreover, some of these even occur deep into the inner regions of the cluster host. The significant number of mergers within clusters is encouraging given the important role expected of mergers in the morphology of galaxies in high density regions.

The final fate of these merger remnants will depend on the mass of the intervening galaxies and their gas content. On the dwarfs regime, mergers are the main mechanism thought to produce blue compact dwarfs (BCDs) with both, observational data as well as idealized numerical simulations showing the feasibility of this formation path  \citep[e.g. ][]{Ostlin_2001, Bekki_1998}. Such mergers will typically involve two dwarf galaxies with similar masses and should contain gas to fuel the central starburst that gives rise to the dense inner blue core. On the other hand, gas-free similar mass dwarf-dwarf mergers would explain better the presence of low surface brightness shells as found in a couple of dwarfs in deep studies of the outskirts of Virgo \citep{Paudel_2017, Zhang_2020}.

\begin{figure}
	\includegraphics[width=\columnwidth]{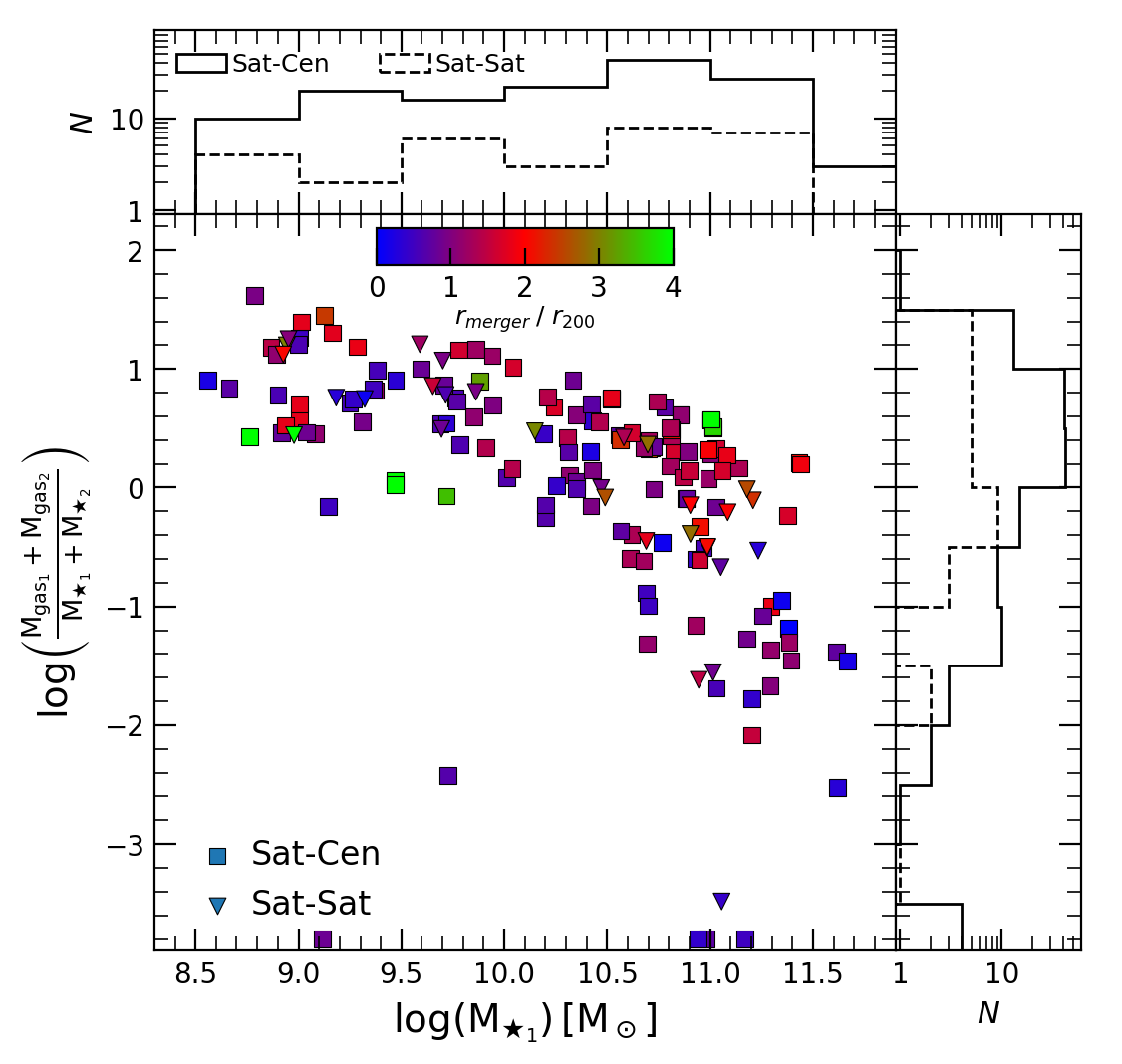}
	\caption{Gas fraction $f_{\rm gas} = M_{\rm gas}/M_{\star}$ in mergers in and around clusters as a function of the stellar mass of the primary galaxy $M_{\star_1}$. Triangle/squares are used for Sat-Sat and Sat-Cen mergers, colour coded according to the location of the merger (see colour bar). Histograms show the distribution along each axis for Sat-Sat and Sat-Cen events. Mergers are gas rich for the dwarfs domain, but transition into a mix of gas rich and gas poor for more massive galaxies. These mergers may provide a natural avenue for the formation of S0s and blue compact dwarf galaxies in galaxy clusters.}
    \label{fig:mergers_fgas}
\end{figure}

For more massive galaxies, mergers could be important contributors to the build up of the S0 (lenticular) population \citep{Baugh_1996, Somerville_1999, Hopkins_2008, Bekki_2008, Arnold_2011}, with some discussion on the need of some gas to re-build the disk \citep{Naab_2006} while other authors find suitable S0 remnants that form even in dry (gas-poor) mergers \citep{Tapia_2014, Eliche-Moral_2018}. However, these numerical simulations have mostly focused on idealized experiments, while the final number of S0-like galaxies expected in clusters from mergers will depend on the location and gas ratios of these mergers occurring within the cosmological set up. 

We explore these scenarios in Fig.~\ref{fig:mergers_fgas}, where we show the gas fraction of the mergers in our clusters (which, as explained above occur all within the galaxy groups) as a function of the primary mass and colour coded by the position within the clusters where the merger occurred. Gas ratios are defined as $f_{\rm gas} = M_{\rm gas}/M_{\star} =  \left(M_{\rm gas_1} + M_{\rm gas_2} \right) / \left(M_{\star_1} + M_{\star_2} \right) $. We find that for dwarf galaxies with $M_{\star} \leq 10^9$\msun\; mergers are in their majority gas rich, supporting this scenario as formation path for blue compact dwarfs in groups and clusters. 

Although several studies of BCDs exist in nearby clusters \citep[e.g. ][]{Zhao_2013, Drinkwater_1996, Vaduvescu_2011, Meyer_2014, Vaduvescu_2014}, very few cover the whole cluster area with completeness enough to quantify the frequency of BCDs compared to other kinds of dwarfs. The Virgo cluster given its relative proximity offers perhaps the best data. Based on photometric plates \citet{Binggeli_1985} reported $\sim 38$ BCDs (member) candidates with B-band magnitude $M_B \leq -13.09$. Of these, \citet{Meyer_2014} studied $30$ objects with $M_r \leq -14$ in detail confirming their BCDs nature and association to Virgo. Assuming a mass-to-light ratio of unity, this sample includes $12$ dwarfs with $M_{\star} \geq 10^{8.2}$\msun, the resolution limit of our study. We find 10 dwarf-dwarf gas-rich mergers that occur within $r<2r_{200}$ in our $10$ clusters sample (considering all galaxies with $M_{\star} \leq 10^9$\msun). This suggests that at least some of these BCDs could be explained by dwarf-dwarf mergers while the majority may have already have fallen in as a BCD. However, more detailed comparisons should be made in the future that include other clusters once complete surveys of BCDs become available in the literature.

Interestingly, a few low-mass galaxies with $M_{\star} <10^{10}$\msun\; experiencing dry mergers are also found ($f_{\rm gas} < 0.1$) in our simulated sample, although they are rare. Whether these events would be enough to explain the observational findings of dwarfs with shell-like features is currently an open question since no complete samples exist of these objects with low surface brightness features. But it is encouraging that within the limited resolution of these simulations a few cases arise that may compare well with these peculiar objects found in Virgo \citep[e.g. ][]{Paudel_2017,Zhang_2020}. 

While most of the mergers in the low-mass end are predicted to be gas-rich, for MW-like galaxies and above we find a wider range of possible gas fractions, in particular with most mergers being gas-poor for $M_{\star} \geq 10^{10.5}$\msun. There is a weak trend suggesting that gas poor mergers occur preferentially within $r_{200}$ of the cluster (blue and purple colours), perhaps highlighting the {\it combined} effects of group infall and cluster environment in removing the gas of $L_{\star}$ galaxies\footnote{Caution might be exercised, as this effect may be affected by the lower gas fractions predicted in groups in Illustris compared to observations, an issue later improved in the related TNG simulations \citep{Pillepich_2018}}. 

We conclude with the remark that mergers can occur within and around galaxy clusters. They are hosted always within the more gentle environment of infalling groups that characterize the hierarchical assembly of clusters. However, we detect an average of $17$ mergers per cluster accumulated during the entire evolution (but up to 42 for our most massive object, 15 of which occur within $r_{200}$). For reference, a cluster like Virgo has $\sim 49$ S0s and another $\sim 30$ dwarfs S0 galaxies \citep{Binggeli_1987,van_den_Bergh_2009}.

This means that mergers, specially for $M_{\star} \geq 10^{10}$\msun\; and above, may have a moderate to significant contribution to the build up of S0s in clusters. However, depending on the particular assembly history and merger events registered in each cluster, it is likely that other mechanisms, for instance the gas removal by ram-pressure \citep{Gunn_Gott_1972} followed by secular effects, will contribute as well to explain the total sample of S0 galaxies today. This is in good agreement with recent observational results suggesting more than one mechanism, including mergers, lead to the formation of lenticular galaxies \citep{Fraser_McKelvie_2018, Coccato_2020, Dolfi_2020}. 

\section{Summary and conclusions}
\label{sec:concl}

We use the Illustris cosmological hydrodynamical simulations to study the assembly of the $10$ most massive galaxy clusters in the box, with typical virial masses $M_{200} \sim 10^{14}$\msun. In particular, we address the contribution of group infall to the surviving population of galaxies, their cluster dynamics and the occurrence of mergers nearby or within clusters. We follow the evolution of galaxies with $M_{\star} \geq 1.5 \times 10^8$\msun, covering a wide range of galaxy masses, from dwarfs comparable to the SMC to the more massive $M_\star \sim 10^{11.5}$\msun\; galaxies in clusters. Our results can be summarized as follows.

\begin{itemize}

\item We find that $38 \pm 15\%$ of present-day galaxies in clusters have fallen in as part of larger groups or galaxy associations, here defined as groups identified by the FoF algorithm. A more strict satellite criteria -- for example requiring that galaxies are within the virial radius of an infalling group -- lowers the result to $\sim 22\%$ of galaxies entering as satellites. This fraction is independent of galaxy stellar mass and shows large dispersion from cluster-to-cluster depending on the particular assembly history. Our results agree well with previous reports in the literature \citep[e.g. ][]{Berrier_2009} and extend the analysis to lower mass dwarfs and also to hydrodynamical simulations compared to N-body only runs. 

\item Poor groups are common: galaxies that are not accreted as single objects will typically come with only one or two companions above our resolution limit. All our clusters accreted at least one group with $10$ galaxies or more, reaching up to $80$ galaxy members in our most extreme case.

\item These groups will quickly evolve in phase-space after infall, making their detection in observations rather difficult. The timescale to double the size and the velocity dispersion of the group is in the range $\tau_d \sim [0.2$-$5$] Gyr, with only a modest dependence on group mass. Instead, different orbits and accretion times are likely to dominate the scatter.
		
\item Although groups become dynamically hotter under the tidal effects of the cluster, the relative velocity between the once group members remains lower than the average of random galaxies in the cluster. This correlated motion, allow for low velocity encounters and, eventually, mergers to occur in the cluster and its surroundings. All identified mergers in our simulation occurred within galaxy groups, with no mergers registered between galaxies that infall as singletons (excluding those mergers with the BCG). 

\item Mergers in groups occur across the entire stellar mass range, from dwarfs to $L_{\star}$ galaxies. The location of these mergers vary, with $\sim 60\%$ occurring outside the clusters, at distances between $1$-$4 r_{200}$, and $\sim 40\%$ happening within the $r_{200}$ of the clusters themselves. Our study offers solid support to the scenario where group infall promotes the occurrence of mergers even within the high velocity dispersion environment of clusters.

\item The gas ratios of these mergers depend critically on galaxy mass: they are strongly gas-rich dominated for $M_{\star} \leq 10^{9}$\msun, while a combination of gas rich and gas poor is found for more massive galaxies with $M_{\star} \geq 10^{10}$\msun. 

\end{itemize}		

Based on studies of idealized (non-cosmological) simulations showing that mergers may contribute to the formation of blue compact dwarfs and S0 galaxies in clusters \citep{Bekki_1998, Bournaud_2005, Eliche-Moral_2018, Bekki_2008}, we conclude that such mergers exist naturally within the $\Lambda$CDM scenario as part of the predicted group infall. The number of events can show large variations from cluster to cluster, but it suggests that mergers may contribute significantly to the build up of the morphological mix even in high density environments. 

We hasten to add that the observed population of S0s in clusters may be larger than the number of predicted mergers in $\Lambda$CDM, requiring of additional formation paths for S0 galaxies. These may include, for instance, secular evolution, gas removal/fading or unstable high redshift disks \citep{van_den_Bergh_2009, Eliche-Moral_2013, Saha_2018} in good agreement with observational constraints. The different kinematical and structural properties of S0s created by these different mechanisms are still challenging to predict with sufficient numerical resolution for simulations in the cosmological context, but observational studies are starting to highlight promising paths to disentangle between the different formation scenarios proposed \citep[see e.g., ][]{Fraser_McKelvie_2018,Coccato_2020}.

Encouragingly, we identify at least two gas-free dwarf-dwarf mergers ($M_{\star} <10^{10}$\msun) in the outskirts of our clusters, which may help explain the discovery of shell-like low-surface brightness stellar features around dwarfs in Virgo. However, higher numerical resolution is needed in order to have more predictive power on the exact morphology of the remnants. Looking forward, efforts like TNG50 \citep{Pillepich_2019, Nelson_2019} or RomulusC \citep{Tremmel_2019} simulations may offer promising new frontiers in our understanding of mergers in cluster environments and their role in the build up of the morphology of cluster galaxies.

\section*{Acknowledgments}

The authors would like to thank the anonymous referee for a constructive and detailed report that helped improve the earlier version of this paper. JAB and MGA acknowledge financial support from CONICET through PIP 11220170100527CO grant. LVS is grateful for funding support from UC-Mexus, NASA-ATP-80NSSC20K0566 and NSF-CAREER-1945310 grants. 

\section*{Data Availability}

This paper is based on halo catalogs and merger trees from the Illustris Project \citep{Vogelsberger_2014a, Vogelsberger_2014b}. These data is publically available at \href{https://www.illustris-project.org/}{https://www.illustris-project.org/}. The catalog of infalling groups, mergers and other products of this analysis may be shared upon request to the corresponding author if no further conflict exists with ongoing projects. 



\bibliographystyle{mnras}
\bibliography{ms.bib} 

\begin{thebibliography}{}
\makeatletter
\relax
\def\mn@urlcharsother{\let\do\@makeother \do\$\do\&\do\#\do\^\do\_\do\%\do\~}
\def\mn@doi{\begingroup\mn@urlcharsother \@ifnextchar [ {\mn@doi@}
  {\mn@doi@[]}}
\def\mn@doi@[#1]#2{\def\@tempa{#1}\ifx\@tempa\@empty \href
  {http://dx.doi.org/#2} {doi:#2}\else \href {http://dx.doi.org/#2} {#1}\fi
  \endgroup}
\def\mn@eprint#1#2{\mn@eprint@#1:#2::\@nil}
\def\mn@eprint@arXiv#1{\href {http://arxiv.org/abs/#1} {{\tt arXiv:#1}}}
\def\mn@eprint@dblp#1{\href {http://dblp.uni-trier.de/rec/bibtex/#1.xml}
  {dblp:#1}}
\def\mn@eprint@#1:#2:#3:#4\@nil{\def\@tempa {#1}\def\@tempb {#2}\def\@tempc
  {#3}\ifx \@tempc \@empty \let \@tempc \@tempb \let \@tempb \@tempa \fi \ifx
  \@tempb \@empty \def\@tempb {arXiv}\fi \@ifundefined
  {mn@eprint@\@tempb}{\@tempb:\@tempc}{\expandafter \expandafter \csname
  mn@eprint@\@tempb\endcsname \expandafter{\@tempc}}}

\bibitem[\protect\citeauthoryear{{Arnold}, {Romanowsky}, {Brodie}, {Chomiuk},
  {Spitler}, {Strader}, {Benson}  \& {Forbes}}{{Arnold}
  et~al.}{2011}]{Arnold_2011}
{Arnold} J.~A.,  {Romanowsky} A.~J.,  {Brodie} J.~P.,  {Chomiuk} L.,  {Spitler}
  L.~R.,  {Strader} J.,  {Benson} A.~J.,   {Forbes} D.~A.,  2011, \mn@doi
  [\apjl] {10.1088/2041-8205/736/2/L26}, \href
  {https://ui.adsabs.harvard.edu/abs/2011ApJ...736L..26A} {736, L26}

\bibitem[\protect\citeauthoryear{{Bah{\'e}} et~al.,}{{Bah{\'e}}
  et~al.}{2019}]{Bahe_2019}
{Bah{\'e}} Y.~M.,  et~al., 2019, \mn@doi [\mnras] {10.1093/mnras/stz361}, \href
  {https://ui.adsabs.harvard.edu/abs/2019MNRAS.485.2287B} {485, 2287}

\bibitem[\protect\citeauthoryear{{Balogh}, {Navarro}  \& {Morris}}{{Balogh}
  et~al.}{2000}]{Balogh_2000}
{Balogh} M.~L.,  {Navarro} J.~F.,   {Morris} S.~L.,  2000, \mn@doi [\apj]
  {10.1086/309323}, \href
  {https://ui.adsabs.harvard.edu/abs/2000ApJ...540..113B} {540, 113}

\bibitem[\protect\citeauthoryear{{Baugh}, {Cole}  \& {Frenk}}{{Baugh}
  et~al.}{1996}]{Baugh_1996}
{Baugh} C.~M.,  {Cole} S.,   {Frenk} C.~S.,  1996, \mn@doi [\mnras]
  {10.1093/mnras/283.4.1361}, \href
  {https://ui.adsabs.harvard.edu/abs/1996MNRAS.283.1361B} {283, 1361}

\bibitem[\protect\citeauthoryear{{Behroozi}, {Wechsler}, {Lu}, {Hahn}, {Busha},
  {Klypin}  \& {Primack}}{{Behroozi} et~al.}{2014}]{Behroozi_2014}
{Behroozi} P.~S.,  {Wechsler} R.~H.,  {Lu} Y.,  {Hahn} O.,  {Busha} M.~T.,
  {Klypin} A.,   {Primack} J.~R.,  2014, \mn@doi [\apj]
  {10.1088/0004-637X/787/2/156}, \href
  {https://ui.adsabs.harvard.edu/abs/2014ApJ...787..156B} {787, 156}

\bibitem[\protect\citeauthoryear{{Bekki}}{{Bekki}}{1998}]{Bekki_1998}
{Bekki} K.,  1998, \mn@doi [\apjl] {10.1086/311508}, \href
  {https://ui.adsabs.harvard.edu/abs/1998ApJ...502L.133B} {502, L133}

\bibitem[\protect\citeauthoryear{{Bekki}}{{Bekki}}{2008}]{Bekki_2008}
{Bekki} K.,  2008, \mn@doi [\mnras] {10.1111/j.1745-3933.2008.00489.x}, \href
  {https://ui.adsabs.harvard.edu/abs/2008MNRAS.388L..10B} {388, L10}

\bibitem[\protect\citeauthoryear{{Berrier}, {Stewart}, {Bullock}, {Purcell},
  {Barton}  \& {Wechsler}}{{Berrier} et~al.}{2009}]{Berrier_2009}
{Berrier} J.~C.,  {Stewart} K.~R.,  {Bullock} J.~S.,  {Purcell} C.~W.,
  {Barton} E.~J.,   {Wechsler} R.~H.,  2009, \mn@doi [\apj]
  {10.1088/0004-637X/690/2/1292}, \href
  {https://ui.adsabs.harvard.edu/abs/2009ApJ...690.1292B} {690, 1292}

\bibitem[\protect\citeauthoryear{{Binggeli}, {Sandage}  \&
  {Tammann}}{{Binggeli} et~al.}{1985}]{Binggeli_1985}
{Binggeli} B.,  {Sandage} A.,   {Tammann} G.~A.,  1985, \mn@doi [\aj]
  {10.1086/113874}, \href
  {https://ui.adsabs.harvard.edu/abs/1985AJ.....90.1681B} {90, 1681}

\bibitem[\protect\citeauthoryear{{Binggeli}, {Tammann}  \&
  {Sandage}}{{Binggeli} et~al.}{1987}]{Binggeli_1987}
{Binggeli} B.,  {Tammann} G.~A.,   {Sandage} A.,  1987, \mn@doi [\aj]
  {10.1086/114467}, \href
  {https://ui.adsabs.harvard.edu/abs/1987AJ.....94..251B} {94, 251}

\bibitem[\protect\citeauthoryear{{Biviano}, {Katgert}, {Thomas}  \&
  {Adami}}{{Biviano} et~al.}{2002}]{Biviano_2002}
{Biviano} A.,  {Katgert} P.,  {Thomas} T.,   {Adami} C.,  2002, \mn@doi [\aap]
  {10.1051/0004-6361:20020340}, \href
  {https://ui.adsabs.harvard.edu/abs/2002A&A...387....8B} {387, 8}

\bibitem[\protect\citeauthoryear{{Bournaud}, {Jog}  \& {Combes}}{{Bournaud}
  et~al.}{2005}]{Bournaud_2005}
{Bournaud} F.,  {Jog} C.~J.,   {Combes} F.,  2005, \mn@doi [\aap]
  {10.1051/0004-6361:20042036}, \href
  {https://ui.adsabs.harvard.edu/abs/2005A&A...437...69B} {437, 69}

\bibitem[\protect\citeauthoryear{{Boylan-Kolchin}, {Springel}, {White}  \&
  {Jenkins}}{{Boylan-Kolchin} et~al.}{2010}]{Boylan-Kolchin_2010}
{Boylan-Kolchin} M.,  {Springel} V.,  {White} S. D.~M.,   {Jenkins} A.,  2010,
  \mn@doi [\mnras] {10.1111/j.1365-2966.2010.16774.x}, \href
  {https://ui.adsabs.harvard.edu/abs/2010MNRAS.406..896B} {406, 896}

\bibitem[\protect\citeauthoryear{{Busha}, {Marshall}, {Wechsler}, {Klypin}  \&
  {Primack}}{{Busha} et~al.}{2011}]{Busha_2011b}
{Busha} M.~T.,  {Marshall} P.~J.,  {Wechsler} R.~H.,  {Klypin} A.,   {Primack}
  J.,  2011, \mn@doi [\apj] {10.1088/0004-637X/743/1/40}, \href
  {https://ui.adsabs.harvard.edu/abs/2011ApJ...743...40B} {743, 40}

\bibitem[\protect\citeauthoryear{{Choque-Challapa}, {Smith}, {Candlish},
  {Peletier}  \& {Shin}}{{Choque-Challapa} et~al.}{2019}]{Choque-Challapa_2019}
{Choque-Challapa} N.,  {Smith} R.,  {Candlish} G.,  {Peletier} R.,   {Shin} J.,
   2019, \mn@doi [\mnras] {10.1093/mnras/stz2829}, \href
  {https://ui.adsabs.harvard.edu/abs/2019MNRAS.490.3654C} {490, 3654}

\bibitem[\protect\citeauthoryear{{Coccato} et~al.,}{{Coccato}
  et~al.}{2020}]{Coccato_2020}
{Coccato} L.,  et~al., 2020, \mn@doi [\mnras] {10.1093/mnras/stz3592}, \href
  {https://ui.adsabs.harvard.edu/abs/2020MNRAS.492.2955C} {492, 2955}

\bibitem[\protect\citeauthoryear{{Cohn}}{{Cohn}}{2012}]{Cohn_2012}
{Cohn} J.~D.,  2012, \mn@doi [\mnras] {10.1111/j.1365-2966.2011.19756.x}, \href
  {https://ui.adsabs.harvard.edu/abs/2012MNRAS.419.1017C} {419, 1017}

\bibitem[\protect\citeauthoryear{{Conselice} \& {Gallagher}}{{Conselice} \&
  {Gallagher}}{1998}]{Conselice_1998}
{Conselice} C.~J.,  {Gallagher} John~S. I.,  1998, \mn@doi [\mnras]
  {10.1046/j.1365-8711.1998.01717.x}, \href
  {https://ui.adsabs.harvard.edu/abs/1998MNRAS.297L..34C} {297, L34}

\bibitem[\protect\citeauthoryear{{D'Onghia}}{{D'Onghia}}{2008}]{Donghia_2008}
{D'Onghia} E.,  2008, arXiv e-prints, \href
  {https://ui.adsabs.harvard.edu/abs/2008arXiv0802.0302D} {p. arXiv:0802.0302}

\bibitem[\protect\citeauthoryear{{D'Onofrio}, {Marziani}  \&
  {Buson}}{{D'Onofrio} et~al.}{2015}]{Donofrio_2015}
{D'Onofrio} M.,  {Marziani} P.,   {Buson} L.,  2015, \mn@doi [Frontiers in
  Astronomy and Space Sciences] {10.3389/fspas.2015.00004}, \href
  {https://ui.adsabs.harvard.edu/abs/2015FrASS...2....4D} {2, 4}

\bibitem[\protect\citeauthoryear{{Davis}, {Efstathiou}, {Frenk}  \&
  {White}}{{Davis} et~al.}{1985}]{Davis_1985}
{Davis} M.,  {Efstathiou} G.,  {Frenk} C.~S.,   {White} S.~D.~M.,  1985,
  \mn@doi [\apj] {10.1086/163168}, \href
  {https://ui.adsabs.harvard.edu/abs/1985ApJ...292..371D} {292, 371}

\bibitem[\protect\citeauthoryear{{De Lucia}}{{De Lucia}}{2012}]{deLucia_2012MW}
{De Lucia} G.,  2012, \mn@doi [Astronomische Nachrichten]
  {10.1002/asna.201211683}, \href
  {https://ui.adsabs.harvard.edu/abs/2012AN....333..460D} {333, 460}

\bibitem[\protect\citeauthoryear{{De Lucia}, {Weinmann}, {Poggianti},
  {Arag{\'o}n-Salamanca}  \& {Zaritsky}}{{De Lucia}
  et~al.}{2012}]{deLucia_2012c}
{De Lucia} G.,  {Weinmann} S.,  {Poggianti} B.~M.,  {Arag{\'o}n-Salamanca} A.,
   {Zaritsky} D.,  2012, \mn@doi [\mnras] {10.1111/j.1365-2966.2012.20983.x},
  \href {https://ui.adsabs.harvard.edu/abs/2012MNRAS.423.1277D} {423, 1277}

\bibitem[\protect\citeauthoryear{{Diemer}, {Mansfield}, {Kravtsov}  \&
  {More}}{{Diemer} et~al.}{2017}]{Diemer_2017}
{Diemer} B.,  {Mansfield} P.,  {Kravtsov} A.~V.,   {More} S.,  2017, \mn@doi
  [\apj] {10.3847/1538-4357/aa79ab}, \href
  {https://ui.adsabs.harvard.edu/abs/2017ApJ...843..140D} {843, 140}

\bibitem[\protect\citeauthoryear{{Dolag}, {Borgani}, {Murante}  \&
  {Springel}}{{Dolag} et~al.}{2009}]{Dolag_2009}
{Dolag} K.,  {Borgani} S.,  {Murante} G.,   {Springel} V.,  2009, \mn@doi
  [\mnras] {10.1111/j.1365-2966.2009.15034.x}, \href
  {https://ui.adsabs.harvard.edu/abs/2009MNRAS.399..497D} {399, 497}

\bibitem[\protect\citeauthoryear{{Dolfi} et~al.,}{{Dolfi}
  et~al.}{2020}]{Dolfi_2020}
{Dolfi} A.,  et~al., 2020, \mn@doi [\mnras] {10.1093/mnras/staa1080}, \href
  {https://ui.adsabs.harvard.edu/abs/2020MNRAS.tmp.1236D} {}

\bibitem[\protect\citeauthoryear{{Drinkwater}, {Currie}, {Young}, {Hardy}  \&
  {Yearsley}}{{Drinkwater} et~al.}{1996}]{Drinkwater_1996}
{Drinkwater} M.~J.,  {Currie} M.~J.,  {Young} C.~K.,  {Hardy} E.,   {Yearsley}
  J.~M.,  1996, \mn@doi [\mnras] {10.1093/mnras/279.2.595}, \href
  {https://ui.adsabs.harvard.edu/abs/1996MNRAS.279..595D} {279, 595}

\bibitem[\protect\citeauthoryear{{Drinkwater}, {Gregg}  \&
  {Colless}}{{Drinkwater} et~al.}{2001}]{Drinkwater_2001}
{Drinkwater} M.~J.,  {Gregg} M.~D.,   {Colless} M.,  2001, \mn@doi [\apjl]
  {10.1086/319113}, \href
  {https://ui.adsabs.harvard.edu/abs/2001ApJ...548L.139D} {548, L139}

\bibitem[\protect\citeauthoryear{{Duffy}, {Schaye}, {Kay}  \& {Dalla
  Vecchia}}{{Duffy} et~al.}{2008}]{Duffy_2008}
{Duffy} A.~R.,  {Schaye} J.,  {Kay} S.~T.,   {Dalla Vecchia} C.,  2008, \mn@doi
  [\mnras] {10.1111/j.1745-3933.2008.00537.x}, \href
  {https://ui.adsabs.harvard.edu/abs/2008MNRAS.390L..64D} {390, L64}

\bibitem[\protect\citeauthoryear{{Eliche-Moral}, {Gonz{\'a}lez-Garc{\'\i}a},
  {Aguerri}, {Gallego}, {Zamorano}, {Balcells}  \& {Prieto}}{{Eliche-Moral}
  et~al.}{2013}]{Eliche-Moral_2013}
{Eliche-Moral} M.~C.,  {Gonz{\'a}lez-Garc{\'\i}a} A.~C.,  {Aguerri} J.~A.~L.,
  {Gallego} J.,  {Zamorano} J.,  {Balcells} M.,   {Prieto} M.,  2013, \mn@doi
  [\aap] {10.1051/0004-6361/201220841}, \href
  {https://ui.adsabs.harvard.edu/abs/2013A&A...552A..67E} {552, A67}

\bibitem[\protect\citeauthoryear{{Eliche-Moral}, {Rodr{\'\i}guez-P{\'e}rez},
  {Borlaff}, {Querejeta}  \& {Tapia}}{{Eliche-Moral}
  et~al.}{2018}]{Eliche-Moral_2018}
{Eliche-Moral} M.~C.,  {Rodr{\'\i}guez-P{\'e}rez} C.,  {Borlaff} A.,
  {Querejeta} M.,   {Tapia} T.,  2018, \mn@doi [\aap]
  {10.1051/0004-6361/201832911}, \href
  {https://ui.adsabs.harvard.edu/abs/2018A&A...617A.113E} {617, A113}

\bibitem[\protect\citeauthoryear{{Fakhouri} \& {Ma}}{{Fakhouri} \&
  {Ma}}{2008}]{Fakhouri_2008}
{Fakhouri} O.,  {Ma} C.-P.,  2008, \mn@doi [\mnras]
  {10.1111/j.1365-2966.2008.13075.x}, \href
  {https://ui.adsabs.harvard.edu/abs/2008MNRAS.386..577F} {386, 577}

\bibitem[\protect\citeauthoryear{{Fakhouri}, {Ma}  \&
  {Boylan-Kolchin}}{{Fakhouri} et~al.}{2010}]{Fakhouri_2010a}
{Fakhouri} O.,  {Ma} C.-P.,   {Boylan-Kolchin} M.,  2010, \mn@doi [\mnras]
  {10.1111/j.1365-2966.2010.16859.x}, \href
  {https://ui.adsabs.harvard.edu/abs/2010MNRAS.406.2267F} {406, 2267}

\bibitem[\protect\citeauthoryear{{Ferrarese} et~al.,}{{Ferrarese}
  et~al.}{2020}]{Ferrarese_2020}
{Ferrarese} L.,  et~al., 2020, \mn@doi [\apj] {10.3847/1538-4357/ab339f}, \href
  {https://ui.adsabs.harvard.edu/abs/2020ApJ...890..128F} {890, 128}

\bibitem[\protect\citeauthoryear{{Fraser-McKelvie}, {Arag{\'o}n-Salamanca},
  {Merrifield}, {Tabor}, {Bernardi}, {Drory}, {Parikh}  \&
  {Argudo-Fern{\'a}ndez}}{{Fraser-McKelvie}
  et~al.}{2018}]{Fraser_McKelvie_2018}
{Fraser-McKelvie} A.,  {Arag{\'o}n-Salamanca} A.,  {Merrifield} M.,  {Tabor}
  M.,  {Bernardi} M.,  {Drory} N.,  {Parikh} T.,   {Argudo-Fern{\'a}ndez} M.,
  2018, \mn@doi [\mnras] {10.1093/mnras/sty2563}, \href
  {https://ui.adsabs.harvard.edu/abs/2018MNRAS.481.5580F} {481, 5580}

\bibitem[\protect\citeauthoryear{{Genel} et~al.,}{{Genel}
  et~al.}{2014}]{Genel_2014}
{Genel} S.,  et~al., 2014, \mn@doi [Monthly Notices of the Royal Astronomical
  Society] {10.1093/mnras/stu1654}, \href
  {https://ui.adsabs.harvard.edu/abs/2014MNRAS.445..175G} {445, 175}

\bibitem[\protect\citeauthoryear{{Giocoli}, {Tormen}  \& {van den
  Bosch}}{{Giocoli} et~al.}{2008}]{Giocoli_2008}
{Giocoli} C.,  {Tormen} G.,   {van den Bosch} F.~C.,  2008, \mn@doi [\mnras]
  {10.1111/j.1365-2966.2008.13182.x}, \href
  {https://ui.adsabs.harvard.edu/abs/2008MNRAS.386.2135G} {386, 2135}

\bibitem[\protect\citeauthoryear{{Giocoli}, {Tormen}, {Sheth}  \& {van den
  Bosch}}{{Giocoli} et~al.}{2010}]{Giocoli_2010}
{Giocoli} C.,  {Tormen} G.,  {Sheth} R.~K.,   {van den Bosch} F.~C.,  2010,
  \mn@doi [\mnras] {10.1111/j.1365-2966.2010.16311.x}, \href
  {https://ui.adsabs.harvard.edu/abs/2010MNRAS.404..502G} {404, 502}

\bibitem[\protect\citeauthoryear{{Gonzalez-Casado}, {Mamon}  \&
  {Salvador-Sole}}{{Gonzalez-Casado} et~al.}{1994}]{Gonzalez-Casado_1994}
{Gonzalez-Casado} G.,  {Mamon} G.~A.,   {Salvador-Sole} E.,  1994, \mn@doi
  [\apjl] {10.1086/187548}, \href
  {https://ui.adsabs.harvard.edu/abs/1994ApJ...433L..61G} {433, L61}

\bibitem[\protect\citeauthoryear{{Gunn} \& {Gott}}{{Gunn} \&
  {Gott}}{1972}]{Gunn_Gott_1972}
{Gunn} J.~E.,  {Gott} J.~Richard I.,  1972, \mn@doi [\apj] {10.1086/151605},
  \href {https://ui.adsabs.harvard.edu/abs/1972ApJ...176....1G} {176, 1}

\bibitem[\protect\citeauthoryear{{Gurzadyan} \& {Mazure}}{{Gurzadyan} \&
  {Mazure}}{1998}]{Gurzadvan_1998}
{Gurzadyan} V.~G.,  {Mazure} A.,  1998, \mn@doi [\mnras]
  {10.1046/j.1365-8711.1998.29511274.x}, \href
  {https://ui.adsabs.harvard.edu/abs/1998MNRAS.295..177G} {295, 177}

\bibitem[\protect\citeauthoryear{{Hinshaw} et~al.,}{{Hinshaw}
  et~al.}{2013}]{Hinshaw_2013}
{Hinshaw} G.,  et~al., 2013, \mn@doi [\apjs] {10.1088/0067-0049/208/2/19},
  \href {https://ui.adsabs.harvard.edu/abs/2013ApJS..208...19H} {208, 19}

\bibitem[\protect\citeauthoryear{{Hopkins}, {Cox}, {Kere{\v{s}}}  \&
  {Hernquist}}{{Hopkins} et~al.}{2008}]{Hopkins_2008}
{Hopkins} P.~F.,  {Cox} T.~J.,  {Kere{\v{s}}} D.,   {Hernquist} L.,  2008,
  \mn@doi [\apjs] {10.1086/524363}, \href
  {https://ui.adsabs.harvard.edu/abs/2008ApJS..175..390H} {175, 390}

\bibitem[\protect\citeauthoryear{{Iodice} et~al.,}{{Iodice}
  et~al.}{2017}]{Iodice_2017}
{Iodice} E.,  et~al., 2017, \mn@doi [\apj] {10.3847/1538-4357/aa6846}, \href
  {https://ui.adsabs.harvard.edu/abs/2017ApJ...839...21I} {839, 21}

\bibitem[\protect\citeauthoryear{{Iodice} et~al.,}{{Iodice}
  et~al.}{2019}]{Iodice_2019}
{Iodice} E.,  et~al., 2019, arXiv e-prints, \href
  {https://ui.adsabs.harvard.edu/abs/2019arXiv190608187I} {p. arXiv:1906.08187}

\bibitem[\protect\citeauthoryear{{Jauzac} et~al.,}{{Jauzac}
  et~al.}{2016}]{Jauzac_2016}
{Jauzac} M.,  et~al., 2016, \mn@doi [\mnras] {10.1093/mnras/stw2251}, \href
  {https://ui.adsabs.harvard.edu/abs/2016MNRAS.463.3876J} {463, 3876}

\bibitem[\protect\citeauthoryear{{Jethwa}, {Erkal}  \& {Belokurov}}{{Jethwa}
  et~al.}{2016}]{Jethwa_2016}
{Jethwa} P.,  {Erkal} D.,   {Belokurov} V.,  2016, \mn@doi [\mnras]
  {10.1093/mnras/stw1343}, \href
  {https://ui.adsabs.harvard.edu/abs/2016MNRAS.461.2212J} {461, 2212}

\bibitem[\protect\citeauthoryear{{Kallivayalil}, {van der Marel}  \&
  {Alcock}}{{Kallivayalil} et~al.}{2006}]{kallivayalil_2006}
{Kallivayalil} N.,  {van der Marel} R.~P.,   {Alcock} C.,  2006, \mn@doi [\apj]
  {10.1086/508014}, \href
  {https://ui.adsabs.harvard.edu/abs/2006ApJ...652.1213K} {652, 1213}

\bibitem[\protect\citeauthoryear{{Kallivayalil}, {van der Marel}, {Besla},
  {Anderson}  \& {Alcock}}{{Kallivayalil} et~al.}{2013}]{kallivayalil_2013}
{Kallivayalil} N.,  {van der Marel} R.~P.,  {Besla} G.,  {Anderson} J.,
  {Alcock} C.,  2013, \mn@doi [\apj] {10.1088/0004-637X/764/2/161}, \href
  {https://ui.adsabs.harvard.edu/abs/2013ApJ...764..161K} {764, 161}

\bibitem[\protect\citeauthoryear{{Karachentsev}, {Tully}, {Wu}, {Shaya}  \&
  {Dolphin}}{{Karachentsev} et~al.}{2014}]{Karachentsev_2014}
{Karachentsev} I.~D.,  {Tully} R.~B.,  {Wu} P.-F.,  {Shaya} E.~J.,   {Dolphin}
  A.~E.,  2014, \mn@doi [\apj] {10.1088/0004-637X/782/1/4}, \href
  {https://ui.adsabs.harvard.edu/abs/2014ApJ...782....4K} {782, 4}

\bibitem[\protect\citeauthoryear{{Knebe}, {Power}, {Gill}  \& {Gibson}}{{Knebe}
  et~al.}{2006}]{Knebe_2006}
{Knebe} A.,  {Power} C.,  {Gill} S. P.~D.,   {Gibson} B.~K.,  2006, \mn@doi
  [\mnras] {10.1111/j.1365-2966.2006.10161.x}, \href
  {https://ui.adsabs.harvard.edu/abs/2006MNRAS.368..741K} {368, 741}

\bibitem[\protect\citeauthoryear{{Lisker}, {Vijayaraghavan}, {Janz},
  {Gallagher}, {Engler}  \& {Urich}}{{Lisker} et~al.}{2018}]{Lisker_2018}
{Lisker} T.,  {Vijayaraghavan} R.,  {Janz} J.,  {Gallagher} John~S. I.,
  {Engler} C.,   {Urich} L.,  2018, \mn@doi [\apj] {10.3847/1538-4357/aadae1},
  \href {https://ui.adsabs.harvard.edu/abs/2018ApJ...865...40L} {865, 40}

\bibitem[\protect\citeauthoryear{{Ludlow}, {Navarro}, {Springel}, {Jenkins},
  {Frenk}  \& {Helmi}}{{Ludlow} et~al.}{2009}]{Ludlow_2009}
{Ludlow} A.~D.,  {Navarro} J.~F.,  {Springel} V.,  {Jenkins} A.,  {Frenk}
  C.~S.,   {Helmi} A.,  2009, \mn@doi [\apj] {10.1088/0004-637X/692/1/931},
  \href {https://ui.adsabs.harvard.edu/abs/2009ApJ...692..931L} {692, 931}

\bibitem[\protect\citeauthoryear{{McGee}, {Balogh}, {Bower}, {Font}  \&
  {McCarthy}}{{McGee} et~al.}{2009}]{McGee_2009}
{McGee} S.~L.,  {Balogh} M.~L.,  {Bower} R.~G.,  {Font} A.~S.,   {McCarthy}
  I.~G.,  2009, \mn@doi [\mnras] {10.1111/j.1365-2966.2009.15507.x}, \href
  {https://ui.adsabs.harvard.edu/abs/2009MNRAS.400..937M} {400, 937}

\bibitem[\protect\citeauthoryear{{Meyer}, {Lisker}, {Janz}  \&
  {Papaderos}}{{Meyer} et~al.}{2014}]{Meyer_2014}
{Meyer} H.~T.,  {Lisker} T.,  {Janz} J.,   {Papaderos} P.,  2014, \mn@doi
  [\aap] {10.1051/0004-6361/201220700}, \href
  {https://ui.adsabs.harvard.edu/abs/2014A&A...562A..49M} {562, A49}

\bibitem[\protect\citeauthoryear{{Mihos}}{{Mihos}}{2003}]{Mihos_2003}
{Mihos} C.,  2003, arXiv e-prints, \href
  {https://ui.adsabs.harvard.edu/abs/2003astro.ph..5512M} {pp
  astro--ph/0305512}

\bibitem[\protect\citeauthoryear{{Moss}}{{Moss}}{2006}]{Moss_2006}
{Moss} C.,  2006, \mn@doi [\mnras] {10.1111/j.1365-2966.2006.11000.x}, \href
  {https://ui.adsabs.harvard.edu/abs/2006MNRAS.373..167M} {373, 167}

\bibitem[\protect\citeauthoryear{{Muriel} \& {Coenda}}{{Muriel} \&
  {Coenda}}{2014}]{Muriel_2014}
{Muriel} H.,  {Coenda} V.,  2014, \mn@doi [\aap] {10.1051/0004-6361/201322033},
  \href {https://ui.adsabs.harvard.edu/abs/2014A&A...564A..85M} {564, A85}

\bibitem[\protect\citeauthoryear{{Naab}, {Jesseit}  \& {Burkert}}{{Naab}
  et~al.}{2006}]{Naab_2006}
{Naab} T.,  {Jesseit} R.,   {Burkert} A.,  2006, \mn@doi [\mnras]
  {10.1111/j.1365-2966.2006.10902.x}, \href
  {https://ui.adsabs.harvard.edu/abs/2006MNRAS.372..839N} {372, 839}

\bibitem[\protect\citeauthoryear{{Natarajan}, {Kneib}, {Smail}, {Treu},
  {Ellis}, {Moran}, {Limousin}  \& {Czoske}}{{Natarajan}
  et~al.}{2009}]{Natarajan_2009}
{Natarajan} P.,  {Kneib} J.-P.,  {Smail} I.,  {Treu} T.,  {Ellis} R.,  {Moran}
  S.,  {Limousin} M.,   {Czoske} O.,  2009, \mn@doi [\apj]
  {10.1088/0004-637X/693/1/970}, \href
  {https://ui.adsabs.harvard.edu/abs/2009ApJ...693..970N} {693, 970}

\bibitem[\protect\citeauthoryear{{Navarro}, {Frenk}  \& {White}}{{Navarro}
  et~al.}{1996}]{Navarro_1996}
{Navarro} J.~F.,  {Frenk} C.~S.,   {White} S. D.~M.,  1996, \mn@doi [\apj]
  {10.1086/177173}, \href
  {https://ui.adsabs.harvard.edu/abs/1996ApJ...462..563N} {462, 563}

\bibitem[\protect\citeauthoryear{{Navarro}, {Frenk}  \& {White}}{{Navarro}
  et~al.}{1997}]{Navarro_1997}
{Navarro} J.~F.,  {Frenk} C.~S.,   {White} S. D.~M.,  1997, \mn@doi [\apj]
  {10.1086/304888}, \href
  {https://ui.adsabs.harvard.edu/abs/1997ApJ...490..493N} {490, 493}

\bibitem[\protect\citeauthoryear{{Nelson} et~al.,}{{Nelson}
  et~al.}{2015}]{Nelson_2015}
{Nelson} D.,  et~al., 2015, \mn@doi [Astronomy and Computing]
  {10.1016/j.ascom.2015.09.003}, \href
  {https://ui.adsabs.harvard.edu/abs/2015A&C....13...12N} {13, 12}

\bibitem[\protect\citeauthoryear{{Nelson} et~al.,}{{Nelson}
  et~al.}{2019}]{Nelson_2019}
{Nelson} D.,  et~al., 2019, \mn@doi [\mnras] {10.1093/mnras/stz2306}, \href
  {https://ui.adsabs.harvard.edu/abs/2019MNRAS.490.3234N} {490, 3234}

\bibitem[\protect\citeauthoryear{{{\"O}stlin}, {Amram}, {Bergvall}, {Masegosa},
  {Boulesteix}  \& {M{\'a}rquez}}{{{\"O}stlin} et~al.}{2001}]{Ostlin_2001}
{{\"O}stlin} G.,  {Amram} P.,  {Bergvall} N.,  {Masegosa} J.,  {Boulesteix} J.,
    {M{\'a}rquez} I.,  2001, \mn@doi [\aap] {10.1051/0004-6361:20010832}, \href
  {https://ui.adsabs.harvard.edu/abs/2001A&A...374..800O} {374, 800}

\bibitem[\protect\citeauthoryear{{Ostriker}}{{Ostriker}}{1980}]{Ostriker_1980}
{Ostriker} J.~P.,  1980, Comments on Astrophysics, \href
  {https://ui.adsabs.harvard.edu/abs/1980ComAp...8..177O} {8, 177}

\bibitem[\protect\citeauthoryear{{Owers}, {Couch}  \& {Nulsen}}{{Owers}
  et~al.}{2009a}]{Owers_2009a}
{Owers} M.~S.,  {Couch} W.~J.,   {Nulsen} P. E.~J.,  2009a, \mn@doi [\apj]
  {10.1088/0004-637X/693/1/901}, \href
  {https://ui.adsabs.harvard.edu/abs/2009ApJ...693..901O} {693, 901}

\bibitem[\protect\citeauthoryear{{Owers}, {Nulsen}, {Couch}  \&
  {Markevitch}}{{Owers} et~al.}{2009b}]{Owers_2009b}
{Owers} M.~S.,  {Nulsen} P. E.~J.,  {Couch} W.~J.,   {Markevitch} M.,  2009b,
  \mn@doi [\apj] {10.1088/0004-637X/704/2/1349}, \href
  {https://ui.adsabs.harvard.edu/abs/2009ApJ...704.1349O} {704, 1349}

\bibitem[\protect\citeauthoryear{{Paudel} et~al.,}{{Paudel}
  et~al.}{2017}]{Paudel_2017}
{Paudel} S.,  et~al., 2017, \mn@doi [\apj] {10.3847/1538-4357/834/1/66}, \href
  {https://ui.adsabs.harvard.edu/abs/2017ApJ...834...66P} {834, 66}

\bibitem[\protect\citeauthoryear{{Pillepich} et~al.,}{{Pillepich}
  et~al.}{2018}]{Pillepich_2018}
{Pillepich} A.,  et~al., 2018, \mn@doi [\mnras] {10.1093/mnras/stx2656}, \href
  {https://ui.adsabs.harvard.edu/abs/2018MNRAS.473.4077P} {473, 4077}

\bibitem[\protect\citeauthoryear{{Pillepich} et~al.,}{{Pillepich}
  et~al.}{2019}]{Pillepich_2019}
{Pillepich} A.,  et~al., 2019, \mn@doi [\mnras] {10.1093/mnras/stz2338}, \href
  {https://ui.adsabs.harvard.edu/abs/2019MNRAS.490.3196P} {490, 3196}

\bibitem[\protect\citeauthoryear{{Pimbblet}}{{Pimbblet}}{2011}]{Pimbblet_2011}
{Pimbblet} K.~A.,  2011, \mn@doi [\mnras] {10.1111/j.1365-2966.2010.17869.x},
  \href {https://ui.adsabs.harvard.edu/abs/2011MNRAS.411.2637P} {411, 2637}

\bibitem[\protect\citeauthoryear{{Rines} \& {Geller}}{{Rines} \&
  {Geller}}{2008}]{Rines_2008}
{Rines} K.,  {Geller} M.~J.,  2008, \mn@doi [\aj]
  {10.1088/0004-6256/135/5/1837}, \href
  {https://ui.adsabs.harvard.edu/abs/2008AJ....135.1837R} {135, 1837}

\bibitem[\protect\citeauthoryear{{Roberts} \& {Parker}}{{Roberts} \&
  {Parker}}{2017}]{Roberts_2017}
{Roberts} I.~D.,  {Parker} L.~C.,  2017, \mn@doi [\mnras]
  {10.1093/mnras/stx317}, \href
  {https://ui.adsabs.harvard.edu/abs/2017MNRAS.467.3268R} {467, 3268}

\bibitem[\protect\citeauthoryear{{Rodriguez-Gomez} et~al.,}{{Rodriguez-Gomez}
  et~al.}{2015}]{Rodriguez-Gomez_2015}
{Rodriguez-Gomez} V.,  et~al., 2015, \mn@doi [\mnras] {10.1093/mnras/stv264},
  \href {https://ui.adsabs.harvard.edu/abs/2015MNRAS.449...49R} {449, 49}

\bibitem[\protect\citeauthoryear{{Rom{\'a}n} \& {Trujillo}}{{Rom{\'a}n} \&
  {Trujillo}}{2017}]{Roman_2017}
{Rom{\'a}n} J.,  {Trujillo} I.,  2017, \mn@doi [\mnras] {10.1093/mnras/stx694},
  \href {https://ui.adsabs.harvard.edu/abs/2017MNRAS.468.4039R} {468, 4039}

\bibitem[\protect\citeauthoryear{{Saha} \& {Cortesi}}{{Saha} \&
  {Cortesi}}{2018}]{Saha_2018}
{Saha} K.,  {Cortesi} A.,  2018, \mn@doi [\apjl] {10.3847/2041-8213/aad23a},
  \href {https://ui.adsabs.harvard.edu/abs/2018ApJ...862L..12S} {862, L12}

\bibitem[\protect\citeauthoryear{{Sales}, {Navarro}, {Abadi}  \&
  {Steinmetz}}{{Sales} et~al.}{2007}]{Sales_2007b}
{Sales} L.~V.,  {Navarro} J.~F.,  {Abadi} M.~G.,   {Steinmetz} M.,  2007,
  \mn@doi [\mnras] {10.1111/j.1365-2966.2007.12024.x}, \href
  {https://ui.adsabs.harvard.edu/abs/2007MNRAS.379.1464S} {379, 1464}

\bibitem[\protect\citeauthoryear{{Sales}, {Navarro}, {Cooper}, {White}, {Frenk}
   \& {Helmi}}{{Sales} et~al.}{2011}]{Sales_2011}
{Sales} L.~V.,  {Navarro} J.~F.,  {Cooper} A.~P.,  {White} S. D.~M.,  {Frenk}
  C.~S.,   {Helmi} A.,  2011, \mn@doi [\mnras]
  {10.1111/j.1365-2966.2011.19514.x}, \href
  {https://ui.adsabs.harvard.edu/abs/2011MNRAS.418..648S} {418, 648}

\bibitem[\protect\citeauthoryear{{Sales}, {Navarro}, {Kallivayalil}  \&
  {Frenk}}{{Sales} et~al.}{2017}]{Sales_2017}
{Sales} L.~V.,  {Navarro} J.~F.,  {Kallivayalil} N.,   {Frenk} C.~S.,  2017,
  \mn@doi [\mnras] {10.1093/mnras/stw2816}, \href
  {https://ui.adsabs.harvard.edu/abs/2017MNRAS.465.1879S} {465, 1879}

\bibitem[\protect\citeauthoryear{{Sarron}, {Adami}, {Durret}  \&
  {Laigle}}{{Sarron} et~al.}{2019}]{Sarron_2019}
{Sarron} F.,  {Adami} C.,  {Durret} F.,   {Laigle} C.,  2019, \mn@doi [\aap]
  {10.1051/0004-6361/201935394}, \href
  {https://ui.adsabs.harvard.edu/abs/2019A&A...632A..49S} {632, A49}

\bibitem[\protect\citeauthoryear{{Sarzi} et~al.,}{{Sarzi}
  et~al.}{2018}]{Sarzi_2018}
{Sarzi} M.,  et~al., 2018, \mn@doi [\aap] {10.1051/0004-6361/201833137}, \href
  {https://ui.adsabs.harvard.edu/abs/2018A&A...616A.121S} {616, A121}

\bibitem[\protect\citeauthoryear{{Sijacki}, {Vogelsberger}, {Genel},
  {Springel}, {Torrey}, {Snyder}, {Nelson}  \& {Hernquist}}{{Sijacki}
  et~al.}{2015}]{Sijacki_2015}
{Sijacki} D.,  {Vogelsberger} M.,  {Genel} S.,  {Springel} V.,  {Torrey} P.,
  {Snyder} G.~F.,  {Nelson} D.,   {Hernquist} L.,  2015, \mn@doi [\mnras]
  {10.1093/mnras/stv1340}, \href
  {https://ui.adsabs.harvard.edu/abs/2015MNRAS.452..575S} {452, 575}

\bibitem[\protect\citeauthoryear{{Somerville} \& {Kolatt}}{{Somerville} \&
  {Kolatt}}{1999}]{Somerville_1999}
{Somerville} R.~S.,  {Kolatt} T.~S.,  1999, \mn@doi [\mnras]
  {10.1046/j.1365-8711.1999.02154.x}, \href
  {https://ui.adsabs.harvard.edu/abs/1999MNRAS.305....1S} {305, 1}

\bibitem[\protect\citeauthoryear{{Springel}}{{Springel}}{2010}]{Springel_2010}
{Springel} V.,  2010, \mn@doi [\mnras] {10.1111/j.1365-2966.2009.15715.x},
  \href {https://ui.adsabs.harvard.edu/abs/2010MNRAS.401..791S} {401, 791}

\bibitem[\protect\citeauthoryear{{Springel}, {White}, {Tormen}  \&
  {Kauffmann}}{{Springel} et~al.}{2001}]{Springel_2001}
{Springel} V.,  {White} S. D.~M.,  {Tormen} G.,   {Kauffmann} G.,  2001,
  \mn@doi [\mnras] {10.1046/j.1365-8711.2001.04912.x}, \href
  {https://ui.adsabs.harvard.edu/abs/2001MNRAS.328..726S} {328, 726}

\bibitem[\protect\citeauthoryear{{Su} et~al.,}{{Su} et~al.}{2019}]{Su_2019}
{Su} Y.,  et~al., 2019, \mn@doi [\aj] {10.3847/1538-3881/ab1d51}, \href
  {https://ui.adsabs.harvard.edu/abs/2019AJ....158....6S} {158, 6}

\bibitem[\protect\citeauthoryear{{Tapia} et~al.,}{{Tapia}
  et~al.}{2014}]{Tapia_2014}
{Tapia} T.,  et~al., 2014, \mn@doi [\aap] {10.1051/0004-6361/201321386}, \href
  {https://ui.adsabs.harvard.edu/abs/2014A&A...565A..31T} {565, A31}

\bibitem[\protect\citeauthoryear{{Taylor} \& {Babul}}{{Taylor} \&
  {Babul}}{2004}]{Taylor_2004}
{Taylor} J.~E.,  {Babul} A.,  2004, \mn@doi [\mnras]
  {10.1111/j.1365-2966.2004.07395.x}, \href
  {https://ui.adsabs.harvard.edu/abs/2004MNRAS.348..811T} {348, 811}

\bibitem[\protect\citeauthoryear{{Taylor} \& {Babul}}{{Taylor} \&
  {Babul}}{2005a}]{Taylor_2005a}
{Taylor} J.~E.,  {Babul} A.,  2005a, \mn@doi [\mnras]
  {10.1111/j.1365-2966.2005.09582.x}, \href
  {https://ui.adsabs.harvard.edu/abs/2005MNRAS.364..515T} {364, 515}

\bibitem[\protect\citeauthoryear{{Taylor} \& {Babul}}{{Taylor} \&
  {Babul}}{2005b}]{Taylor_2005b}
{Taylor} J.~E.,  {Babul} A.,  2005b, \mn@doi [\mnras]
  {10.1111/j.1365-2966.2005.09581.x}, \href
  {https://ui.adsabs.harvard.edu/abs/2005MNRAS.364..535T} {364, 535}

\bibitem[\protect\citeauthoryear{{Tremmel} et~al.,}{{Tremmel}
  et~al.}{2019}]{Tremmel_2019}
{Tremmel} M.,  et~al., 2019, \mn@doi [\mnras] {10.1093/mnras/sty3336}, \href
  {https://ui.adsabs.harvard.edu/abs/2019MNRAS.483.3336T} {483, 3336}

\bibitem[\protect\citeauthoryear{{Trentham} \& {Hodgkin}}{{Trentham} \&
  {Hodgkin}}{2002}]{Trentham_2002}
{Trentham} N.,  {Hodgkin} S.,  2002, \mn@doi [\mnras]
  {10.1046/j.1365-8711.2002.05440.x}, \href
  {https://ui.adsabs.harvard.edu/abs/2002MNRAS.333..423T} {333, 423}

\bibitem[\protect\citeauthoryear{{Treu}, {Ellis}, {Kneib}, {Dressler}, {Smail},
  {Czoske}, {Oemler}  \& {Natarajan}}{{Treu} et~al.}{2003}]{Treu_2003}
{Treu} T.,  {Ellis} R.~S.,  {Kneib} J.-P.,  {Dressler} A.,  {Smail} I.,
  {Czoske} O.,  {Oemler} A.,   {Natarajan} P.,  2003, \mn@doi [\apj]
  {10.1086/375314}, \href
  {https://ui.adsabs.harvard.edu/abs/2003ApJ...591...53T} {591, 53}

\bibitem[\protect\citeauthoryear{{Vaduvescu}, {Kehrig}, {Vilchez}  \&
  {Unda-Sanzana}}{{Vaduvescu} et~al.}{2011}]{Vaduvescu_2011}
{Vaduvescu} O.,  {Kehrig} C.,  {Vilchez} J.~M.,   {Unda-Sanzana} E.,  2011,
  \mn@doi [\aap] {10.1051/0004-6361/201116651}, \href
  {https://ui.adsabs.harvard.edu/abs/2011A&A...533A..65V} {533, A65}

\bibitem[\protect\citeauthoryear{{Vaduvescu}, {Kehrig}, {Bassino}, {Smith
  Castelli}  \& {Calder{\'o}n}}{{Vaduvescu} et~al.}{2014}]{Vaduvescu_2014}
{Vaduvescu} O.,  {Kehrig} C.,  {Bassino} L.~P.,  {Smith Castelli} A.~V.,
  {Calder{\'o}n} J.~P.,  2014, \mn@doi [\aap] {10.1051/0004-6361/201322615},
  \href {https://ui.adsabs.harvard.edu/abs/2014A&A...563A.118V} {563, A118}

\bibitem[\protect\citeauthoryear{{Venhola} et~al.,}{{Venhola}
  et~al.}{2017}]{Venhola_2017}
{Venhola} A.,  et~al., 2017, \mn@doi [\aap] {10.1051/0004-6361/201730696},
  \href {https://ui.adsabs.harvard.edu/abs/2017A&A...608A.142V} {608, A142}

\bibitem[\protect\citeauthoryear{{Vijayaraghavan} \& {Ricker}}{{Vijayaraghavan}
  \& {Ricker}}{2013}]{Vijayaraghavan_2013}
{Vijayaraghavan} R.,  {Ricker} P.~M.,  2013, \mn@doi [\mnras]
  {10.1093/mnras/stt1485}, \href
  {https://ui.adsabs.harvard.edu/abs/2013MNRAS.435.2713V} {435, 2713}

\bibitem[\protect\citeauthoryear{{Vijayaraghavan}, {Gallagher}  \&
  {Ricker}}{{Vijayaraghavan} et~al.}{2015}]{Vijayaraghavan_2015}
{Vijayaraghavan} R.,  {Gallagher} J.~S.,   {Ricker} P.~M.,  2015, \mn@doi
  [\mnras] {10.1093/mnras/stu2761}, \href
  {https://ui.adsabs.harvard.edu/abs/2015MNRAS.447.3623V} {447, 3623}

\bibitem[\protect\citeauthoryear{{Vogelsberger}, {Genel}, {Sijacki}, {Torrey},
  {Springel}  \& {Hernquist}}{{Vogelsberger} et~al.}{2013}]{Vogelsberger_2013}
{Vogelsberger} M.,  {Genel} S.,  {Sijacki} D.,  {Torrey} P.,  {Springel} V.,
  {Hernquist} L.,  2013, \mn@doi [\mnras] {10.1093/mnras/stt1789}, \href
  {https://ui.adsabs.harvard.edu/abs/2013MNRAS.436.3031V} {436, 3031}

\bibitem[\protect\citeauthoryear{{Vogelsberger} et~al.,}{{Vogelsberger}
  et~al.}{2014a}]{Vogelsberger_2014a}
{Vogelsberger} M.,  et~al., 2014a, \mn@doi [Monthly Notices of the Royal
  Astronomical Society] {10.1093/mnras/stu1536}, \href
  {https://ui.adsabs.harvard.edu/abs/2014MNRAS.444.1518V} {444, 1518}

\bibitem[\protect\citeauthoryear{{Vogelsberger} et~al.,}{{Vogelsberger}
  et~al.}{2014b}]{Vogelsberger_2014b}
{Vogelsberger} M.,  et~al., 2014b, \mn@doi [\nat] {10.1038/nature13316}, \href
  {https://ui.adsabs.harvard.edu/abs/2014Natur.509..177V} {509, 177}

\bibitem[\protect\citeauthoryear{{Yang}, {Mo}  \& {van den Bosch}}{{Yang}
  et~al.}{2009}]{Yang_2009}
{Yang} X.,  {Mo} H.~J.,   {van den Bosch} F.~C.,  2009, \mn@doi [\apj]
  {10.1088/0004-637X/695/2/900}, \href
  {https://ui.adsabs.harvard.edu/abs/2009ApJ...695..900Y} {695, 900}

\bibitem[\protect\citeauthoryear{{Yang}, {Mo}, {Zhang}  \& {van den
  Bosch}}{{Yang} et~al.}{2011}]{Yang_2011}
{Yang} X.,  {Mo} H.~J.,  {Zhang} Y.,   {van den Bosch} F.~C.,  2011, \mn@doi
  [\apj] {10.1088/0004-637X/741/1/13}, \href
  {https://ui.adsabs.harvard.edu/abs/2011ApJ...741...13Y} {741, 13}

\bibitem[\protect\citeauthoryear{{Zhang}, {Reiprich}, {Finoguenov}, {Hudson}
  \& {Sarazin}}{{Zhang} et~al.}{2009}]{Zhang_2009}
{Zhang} Y.-Y.,  {Reiprich} T.~H.,  {Finoguenov} A.,  {Hudson} D.~S.,
  {Sarazin} C.~L.,  2009, \mn@doi [\apj] {10.1088/0004-637X/699/2/1178}, \href
  {https://ui.adsabs.harvard.edu/abs/2009ApJ...699.1178Z} {699, 1178}

\bibitem[\protect\citeauthoryear{{Zhang} et~al.,}{{Zhang}
  et~al.}{2020}]{Zhang_2020}
{Zhang} H.-X.,  et~al., 2020, arXiv e-prints, \href
  {https://ui.adsabs.harvard.edu/abs/2020arXiv200209517Z} {p. arXiv:2002.09517}

\bibitem[\protect\citeauthoryear{{Zhao}, {Gao}  \& {Gu}}{{Zhao}
  et~al.}{2013}]{Zhao_2013}
{Zhao} Y.,  {Gao} Y.,   {Gu} Q.,  2013, \mn@doi [\apj]
  {10.1088/0004-637X/764/1/44}, \href
  {https://ui.adsabs.harvard.edu/abs/2013ApJ...764...44Z} {764, 44}

\bibitem[\protect\citeauthoryear{{van den Bergh}}{{van den
  Bergh}}{2009}]{van_den_Bergh_2009}
{van den Bergh} S.,  2009, \mn@doi [\apj] {10.1088/0004-637X/702/2/1502}, \href
  {https://ui.adsabs.harvard.edu/abs/2009ApJ...702.1502V} {702, 1502}

\bibitem[\protect\citeauthoryear{{van den Bosch}, {Tormen}  \& {Giocoli}}{{van
  den Bosch} et~al.}{2005}]{vandenBosch_2005}
{van den Bosch} F.~C.,  {Tormen} G.,   {Giocoli} C.,  2005, \mn@doi [\mnras]
  {10.1111/j.1365-2966.2005.08964.x}, \href
  {https://ui.adsabs.harvard.edu/abs/2005MNRAS.359.1029V} {359, 1029}

\makeatother
\end{thebibliography}



\appendix

\section{Measurement of group size}
\label{sec:appendix}

In the main body of the paper we characterize the timescale of group disruption, $\tau_d$, defined as the time taken for a group to double the r.m.s distance of the member galaxies, $\sigma_r$, compared to their value at infall. We have experimented with several other definitions to ensure that $\tau_d$ is a good measure of the disruption timescale. 

\begin{figure}
	\includegraphics[width=\columnwidth]{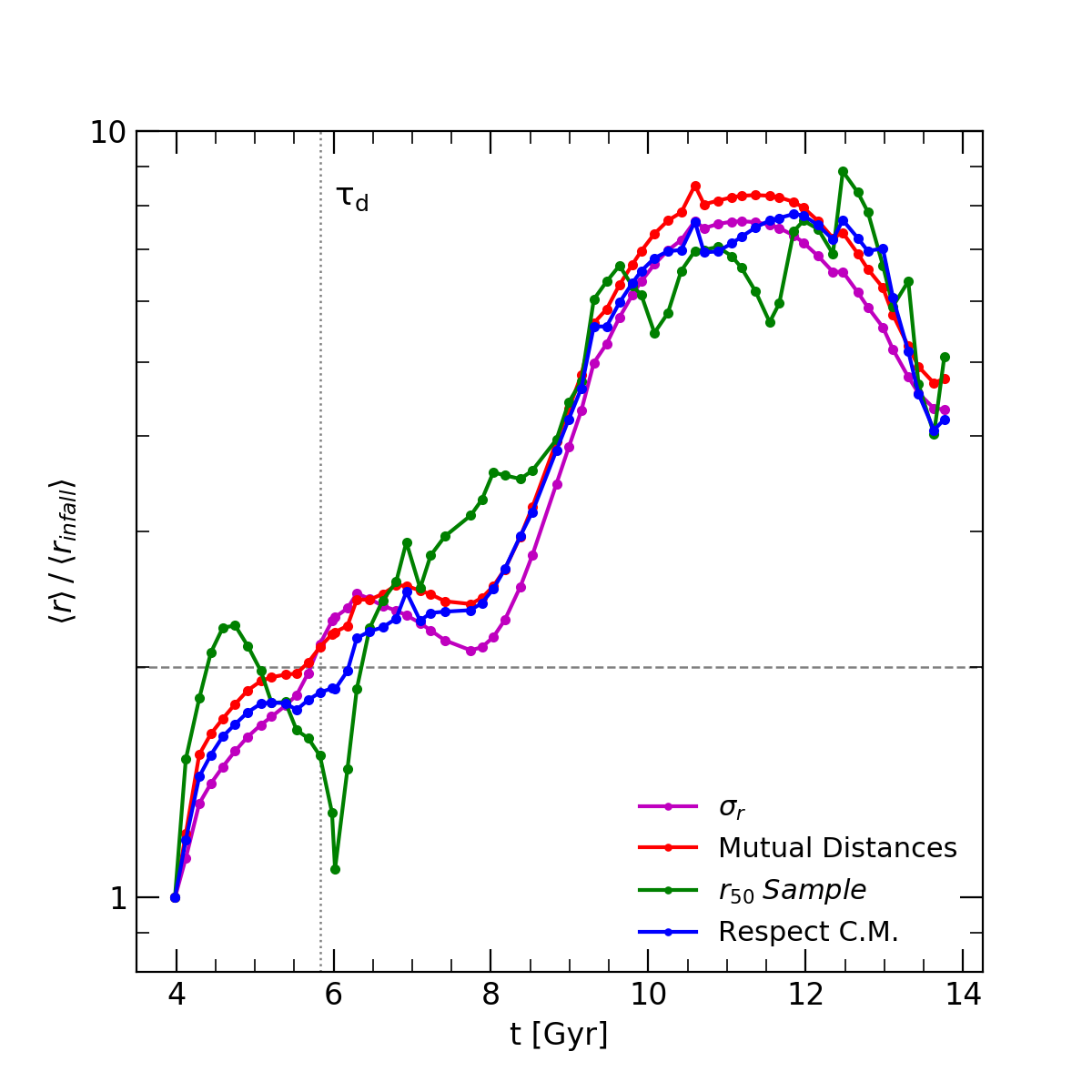}
    \caption{Size evolution with time for one of our infalling groups while it orbits within the host cluster. This group is the same as shown in Fig.~\ref{fig:orbits}, lines start at the identified infall time. Different curves correspond to different definition of size (see text for more detail), with magenta being our definition adopted in the paper (the r.m.s distance between galaxies in the group). Curves have been individually normalized to their infall value. All methods give a similar behavior, with $r_{50}$ (the radius containing half of the group members at different times, shown in green) being slightly more noisy than the rest. Note that the timescale of disruption $\tau_d$, here defined as the time where $\sigma_r$ has doubled with respect to the infall value (dotted vertical line), would be nearly the same regardless of the specific method chosen to quantify the group size.}
    \label{fig:appendix_disruption}
\end{figure}

Fig.~\ref{fig:appendix_disruption} shows for the example group in Fig.~\ref{fig:orbits}, the time evolution of a characteristic (spatial) size defined in several ways: $(i)$ $\sigma_r$ as adopted in the paper (magenta), $(ii)$ the average of distances between galaxy members in the group (red), $(iii)$ the radius enclosing half of the group members $r_{50}$ (green) or $(iv)$ average distances to the center of mass of the group (blue). All values are shown normalized to its infall value. 

As indicated in the figure, all definitions lay very close to each other, a behavior that is common to all our groups analyzed here. More specifically, we use the time $\tau_d$ measured as the time when $\sigma_r$ has doubled its initial value, highlighted by the dotted vertical line, to characterize the spatial disruption timescale. Any of these methods would have resulted on a very similar time measurement. We therefore adopt $\tau_d$ in the remaining of our analysis.

\bsp	
\label{lastpage}
\end{document}